\documentclass[12pt]{article}
\usepackage{amsmath}
\usepackage{cite}

\usepackage{epsfig}

\def\lrang#1{\left\langle#1\right\rangle}
\def\abs#1{\left| \: #1 \: \right|}
\def\la{\mathrel{\mathpalette\fun <}}
\def\ga{\mathrel{\mathpalette\fun >}}
\def\fun#1#2{\lower3.6pt\vbox{\baselineskip0pt\lineskip.9pt
  \ialign{$\mathsurround=0pt#1\hfil##\hfil$\crcr#2\crcr\sim\crcr}}}
\def\cO#1{{\cal{O}}\left(#1\right)}
\def\Im{\mathop{\rm Im}}
\def\Re{\mathop{\rm Re}}

\def\vkti#1{\vec{k}_{t#1}}

\def\tkt{\kappa_t}
\def\tkti#1{\kappa_{t#1}}

\def\vtkt{\vec{\kappa}_t}
\def\vtkti#1{\vec{\kappa}_{t#1}}

\def\tpt{p}
\def\vtpt{\vec{\tpt}}

\def\vb{\vec{b}}

\def\bP{\bar{P}}
\def\Qbar{{\bar Q}}
\def\bmu{\bar \mu}
\def\tC{\tilde{D}}

\def\al{\alpha}
\def\be{\beta}
\def\gam{\gamma}
\def\lam{\lambda}

\def\Om{\Omega}

\def\cA{{\cal{A}}}    
\def\bB{{\cal{B}}}    
\def\cM{{\cal{M}}}    
\def\cP{{\cal{P}}}    

\def\eps{\cP}

\def\cR{{\cal{R}}}               
\def\cRt{{\mathop{\rm\bf R}}}    

\def\half{\mbox{\small $\frac{1}{2}$}}
\def\tq{\mbox{\small $\frac{3}{4}$}}
\def\th{\mbox{\small $\frac{3}{2}$}}
\def\os{\mbox{\small $\frac{1}{6}$}}

\def\PT{\mbox{\scriptsize PT}}
\def\NP{\mbox{\scriptsize NP}}
\def\MSbar{\overline{\mbox{\scriptsize MS}}}
\def\CMW{\mbox{\scriptsize CMW}}

\def\cf{C_F}
\def\ca{C_A}
\def\as{\alpha_{\mbox{\scriptsize s}}}
\def\asb{{\bar \alpha}_{\mbox{\scriptsize s}}}
\def\ae{\alpha_{\mbox{\scriptsize eff}}}
\def\ascmw{\alpha_{\CMW}}
\def\LQCD{\Lambda_{\mbox{\scriptsize QCD}}}

\def\jhep#1#2#3{{JHEP} {\underline {#1}} (19#3) #2}
\def\npb#1#2#3{{ Nucl. Phys.} {\underline {B#1}} (19#3) #2}
\def\plb#1#2#3{{ Phys. Lett.} {\underline {#1B}} (19#3) #2}

\def\pr#1#2#3{{ Phys. Rev.} {\underline {D#1}} (19#3) #2}

\def\zp#1#2#3{{ Zeit.\ Phys.} {\underline {C#1}} (19#3) #2}

\def\epj#1#2#3{{ Eur. Phys. J. } {\underline {C#1}} (19#3) #2}

\numberwithin{equation}{section}

\begin{document}
\begin{titlepage}
\renewcommand{\thefootnote}{\fnsymbol{footnote}}
\begin{flushright}
     Bicocca--FT--98--01 \\
     hep-ph/9812487 \\
     December 1998 \\
\end{flushright}
\par \vskip 10mm
\begin{center}
  {\Large \bf Revisiting non-perturbative effects in the jet
    broadenings\footnote{This work was supported in part by the EU
      Fourth Framework Programme `Training and Mobility of
      Researchers', Network `Quantum Chromodynamics and the Deep
      Structure of Elementary Particles', contract FMRX-CT98-0194 (DG
      12-MIHT).}}
\end{center}
\par \vskip 2mm
\begin{center}
{\bf 
Yu.L. \ Dokshitzer\footnote{
On leave from St. Petersburg Nuclear Institute,
Gatchina, St. Petersburg 188350, Russia},
 G.\ Marchesini, G.P.\ Salam}\\
\vskip 3 mm
Dipartimento di Fisica, Universit\`a di Milano-Bicocca \\
and INFN, Sezione di Milano, Italy
\end{center}

\par \vskip 2mm
\begin{center} {\large \bf Abstract} 
\end{center}
We show that taking into account the interplay between perturbative
and non-perturbative effects, the power-suppressed shift to the
broadening distributions becomes $B$ dependent, and the
non-perturbative contribution to the mean values becomes proportional
to $1/(Q\sqrt{\as(Q)})$. The new theoretical treatment greatly
improves the consistency of the phenomenology with the notion of the
universality of confinement effects in jet shapes.
\begin{quote}
\end{quote}

\end{titlepage}

\section{Introduction}
There is mounting evidence in favour of the universal pattern of
leading $1/Q$ non-perturbative power corrections to collinear/infrared
safe (CIS) observables~\cite{universal}.  They include jet-shape
observables such as the thrust $T$, the $C$-parameter, squared jet
masses and the jet broadening $B$ (definitions can be found, for
example, in \cite{DLMSuniv}).  $1/Q$ power effects are expected
also in the energy-energy correlation function and $\sigma_L$ in
$e^+e^-$ annihilation.  The $1/Q$ power terms are seen both in the
means and the distributions.  For instance, the distribution of events
in $T$ in the 2-jet region $1\gg 1-T \gg \Lambda/Q$ was
predicted~\cite{DokWeb97,KS} to be given by a simple shift of the
corresponding {\em perturbative}\/ QCD prediction by a constant value
inversely proportional to $Q$.  The same pattern was expected to hold
for other jet shapes $V$ as well, with the true distribution in the
kinematical region $1\gg V \gg \Lambda/Q$ being related to its
perturbative ($\PT$) counterpart by a shift,
\begin{equation}
  \label{eq:shift}
  \frac{d\sigma}{dV}(V)\>=\> \frac{d\sigma^{(\PT)}}{dV}(V-
  \Delta_V)\>,
\quad \Delta_V= c_V\cP\>,
\end{equation}
with $c_V$ an observable dependent, perturbatively calculable,
numerical coefficient which for the thrust, $C$-parameter and 
total and heavy-jet masses, e.g., is
\begin{equation}
\mbox{
\begin{minipage}{4in}
\begin{tabular}{|l|c||c||c|c|} \hline 
 $V=$  & $ 1\!-\!T$ & $C$ & $M_T^2/Q^2$ & $M^2_H/Q^2$  \\ 
\hline
  $c_V=$  & 2 & $3\pi$ & 2 & 1  \\
\hline  
\end{tabular}
\end{minipage}
}
\end{equation}
The parameter $\cP\propto 1/Q$ is the non-perturbative ($\NP$)
quantity which effectively ``measures'' the intensity of the QCD
interaction over the infrared momentum region.  Its magnitude can be
interpreted as being related to a mean value of the QCD coupling over
the infrared region, say, $k\le \mu_I=2$~GeV,
\begin{equation}
  \label{eq:a0def}
\al_0(\mu_I) = \mu_I^{-1}\int_0^{\mu_I} dk\> \as(k)\>.
\end{equation}
A proper definition of $\cP$ includes an infrared matching scale
$\mu_I$ which is necessary for merging, in a renormalon-free manner,
the $\PT$ and $\NP$ contributions to a given observable. It also calls
for a two-loop analysis of the non-perturbative
contribution~\cite{DLMSthrust}, 
without which the magnitude of the power correction cannot be
quantified better than up to a factor of order unity.

Experimentally, \NP\  %
effects in the thrust distribution have been found to be consistent
with the shift rule \eqref{eq:shift}, with $\al_0\simeq0.5$. The same
value was experimentally extracted from the $Q$-dependence of
$\lrang{1-T}$ \cite{ALEPH,DELPHI,JADE98,H1}. The $C$-parameter has
also been studied (both the distribution and the mean) and found to be
consistent, with a similar value of $\al_0$ \cite{JADE98}.

Power effects in the broadening $B$ were also expected to shift the
distribution, but with $\Delta_B$ logarithmically enhanced ($\Delta_B
\propto \ln Q/Q$) \cite{DLMSuniv}.  The H1 collaboration however stated
that the data was not consistent with a $\ln Q$ enhancement \cite{H1}.
Most recently the JADE collaboration studied the discrepancy and
showed that relative to the perturbative distribution, the
experimental distribution is not only shifted, but also ``squeezed''
\cite{JADEmontp}.

In this paper we revisit the broadening.  We show that the coefficient
of the power correction shift, $Q\Delta_B$, is neither proportional to
$\ln Q$, nor a constant, but rather is a function logarithmically
depending on $B$.  The reason is the
following.  The non-perturbative
contribution to $V$ comes from the emission of {\em gluers}\/ (gluons with
{\em finite}\/ transverse momenta, of the order of the QCD scale
$\Lambda$, with respect to the quark direction \cite{gluer}).  For
instance, in the soft approximation, the contributions to the thrust,
$C$-parameter and broadening from a secondary parton $i$ can be
expressed as
\begin{equation}
(1-T)_i = \frac{k_{ti}}{Q}\,e^{-\abs{\eta_i}}\>, \qquad
C_i = \frac{3k_{ti}}{Q\cosh{\eta_i}} \>, \qquad
2B_i = \frac{k_{ti}}{Q}\>,
\end{equation}
where the transverse momentum $k_{ti}$ and the rapidity $\eta_i$ are
measured with respect to the thrust axis.  The feature that $1-T$ and
$C$ have in common is that the dominant non-perturbative contribution
to these and similar shapes is determined by the radiation of soft
gluers at {\em large}\/ angles.  This radiation is insensitive to a
tiny mismatch, $\lrang{\Theta_q}=\cO{\as}$, between the quark and
thrust axis directions which is due to perturbative gluon radiation.
Therefore the quark momentum direction can be identified with the
thrust axis in the \PT\ %
and \NP\ %
analysis of $T$ and $C$.

The broadening, on the contrary, accumulates contributions which do not
depend on rapidity, so that the mismatch between the quark and the
thrust axis could be important. As we shall see later, for the $\PT$
analysis of broadening this mismatch is irrelevant, to next-to-leading
accuracy.  However, the mismatch plays a crucial r\^ole for the $\NP$
effects in the broadening, both in the means and the distributions.
 
If one naively assumes that the quark direction can be approximated by
that of the thrust axis, as is the case for the perturbative
contribution, then $B$ accumulates \NP\ %
contributions from gluons with rapidities up to the kinematically
allowed value $\eta_i\le \eta_{\max}\simeq\ln(Q/k_{ti})$.  In this
case one would find the shift in the $B$ spectrum to be
logarithmically enhanced,
\begin{equation}
  \label{old}
\Delta_B \>=\> c_B\cP \cdot \ln \frac{Q}{Q_B}\>,
\end{equation}
where $c_B=1(\half)$ for the total (single-jet; wide-jet)
broadening~\cite{DLMSuniv}.  What this overlooks is the fact that the
{\em uniform}\/ distribution in $\eta_i$ (defined with respect to the
thrust axis) holds only for gluon rapidities not exceeding $|\ln
\Theta_q|$.
{\em Hard}\/ gluons with energies $k_{0i}>k_{ti}/\Theta_q$ are collinear 
to the {\em quark}\/ direction rather than to that of the thrust axis        
and therefore do not contribute essentially to $B$.
As a result, for a given quark angle, the \NP\ %
contribution $\delta B$ to the broadening $B$ comes out proportional
to the quark rapidity (we note the distinction
between the \NP\ %
contribution $\delta B$ to a given
perturbative configuration, and the \NP\ %
shift $\Delta_B$ to the perturbative distribution, where the latter is
integrated over all perturbative configurations leading to
particular value of $B$).
For the single-jet broadening one has
\begin{equation}
\label{eq:nptheta}
  \delta B   \>\simeq\> c_1\cP\cdot \ln\frac1{\Theta_q}\>.
\end{equation}
Let us describe, semi-quantitatively,  
how \eqref{eq:nptheta} affects the $\NP$ corrections to
$\lrang{B}$ and to the $B$ distribution.
The power correction to the mean single-jet broadening
$\lrang{B}_1$ is obtained by evaluating the perturbative average of 
$\delta B_1$ in \eqref{eq:nptheta},
\begin{equation}
  \lrang{B}_1^{(\NP)} \equiv \lrang{B}_1- \lrang{B}_1^{(\PT)}  
  \>\simeq\> c_1\cP\cdot \lrang{\ln\frac1{\Theta_q}}\>.
\end{equation}
The $\PT$-distribution in $\Theta_q$ at the Born level is singular at
$\Theta_q\!=\!0$. In high orders this singularity is damped by the
double-logarithmic Sudakov form factor. 
As a result, the $\NP$ component of $\lrang{B}_1$ gets enhanced by  
\begin{equation}
   \lrang{\ln\frac1{\Theta_q}} \>\simeq\> \frac{\pi}{2\sqrt{C_F\as(Q)}}\>. 
\end{equation}
For the mean wide-jet broadening $\lrang{B}_W$
the result has the same structure
with the replacement $C_F\to 2C_F$ due to the fact that now it is
radiation off {\em two}\/ jets which determines the $\Theta_q$ distribution.

The shift in the single-jet (or wide-jet) broadening can be expressed
as
\begin{equation}
  \Delta_1(B)   \>\simeq\> c_1\cP\cdot\lrang{\ln\frac1{\Theta_q}}_B\>,
\end{equation}
where the average is taken over the perturbative distribution in 
the quark angle $\Theta_q$ while keeping the value of $B$ fixed.
Since $\Theta_q$  is kinematically proportional to $B$,
the $\log$-enhancement of the shift in the $B$ spectrum becomes
\begin{equation}
\label{new}
    \Delta_1(B) \>\simeq\> c_1\cP\cdot\ln\frac{B_0}{B}\>,
\end{equation}
with $B_0=B_0(\as\ln B)$ a calculable function 
slowly depending on $B$. 
Thus, the shift in the $B_1$ ($B_W$) distribution becomes
logarithmically dependent on $B$.

Finally, the shift in the total two-jet broadening distribution
$\Delta_T(B)$ will also be derived. It has a somewhat more complicated
$B$ dependence.  In the kinematical region where the multiplicity of
gluon radiation is small, $\as\ln^2B\ll1$, one of the two jets is
responsible for the whole \PT-component of the event broadening,
while the second is ``empty''. That ``empty'' jet contributes the most
to the shift: in the absence of perturbative radiation the direction
of the quark momentum in this jet stays closer to the thrust axis.
This results in
\begin{equation}
  \Delta_T(B) \>\simeq\> \Delta_1(B) + \lrang{B}_1^{(\NP)} \>\simeq\>
  c_1\cP\left( \ln \frac1{B} +  \frac{\pi}{2\sqrt{C_F\as}}\,
\>+\>\cO{1}\right).
\end{equation}
In these circumstances the $B$ dependence of the total shift
practically coincides with that of a single jet.  In the opposite
regime of well developed \PT\ %
radiation, $\as\ln^2B\gg1$, the jets are forced to share $B$ equally,
and we have
\begin{equation}
   \Delta_T(B) \>\simeq\> 2\cdot \Delta_1(B/2) \>\simeq\> 
2\cdot c_1\cP\> \ln \frac1{B} \>.
\end{equation}

In the present paper we derive the correspondingly modified
predictions for the $B$ {\em dependent}\/ $1/Q$ shifts in the
broadening spectra and the $1/Q$ corrections to the means, which
supersede the earlier results of~\cite{DLMSuniv}.  The paper is
organised as follows.

Section 2 introduces the quantities to be studied, and recollects the
next-to-leading-logarithmic perturbative result for the single-jet,
wide-jet and total distributions.

In section 3 we derive the new results for the power corrections to
the means and distributions.

Section 4 presents a comparison of these new results with experimental 
data.

In the concluding section \ref{sec:Conclusions} 
we discuss our results in the context of the interplay
between $\PT$ and $\NP$ effects.

The remainder of the paper consists of appendices which contain
technical details of the derivation and a Monte Carlo study which
illustrates some important features of the analytical results.
In the last Appendix~\ref{sec:Collection}
we give the full list of formulas to be used to describe
the leading power correction to the broadening means and
distributions.

\section{Broadening distribution}

\subsection{Distributions and means}

Broadening in $e^+e^-$ annihilation with c.m.s.\ energy $Q$ is defined 
as a sum of transverse momenta of final particles with respect to the
thrust axis, 
\begin{equation}
  \label{eq:Bdef}
  2Q\cdot B_{X} \>=\> \sum_{i\in X} \abs{\vec{p}_{\perp i}}\>.
\end{equation}
The symbol $X$ here marks the set of selected final particles. The
three options under discussion are 
\begin{enumerate}
\item single-jet broadening, $B_1$, with the sum running 
      over particles in one hemisphere (right, $B_1=B_R$,  
      or left, $B_1=B_L$), 
\item total broadening of the event $B_T=B_R+B_L$, and 
\item wide-jet broadening which is the larger of the two, 
      $B_W=\max\{B_R, B_L\}$ on an event by event basis. 
\end{enumerate}
We define the integrated single-jet distribution as
\begin{equation}
   \Sigma_1(B) = \int_0^{B} dB\>  \sigma^{-1}\frac{d\sigma}{dB} 
 \>, \quad \Sigma_1'(B) \equiv  \frac{d\Sigma_1}{dB}  \>=\>
\sigma^{-1}\frac{d\sigma}{dB}\>.
\end{equation}
For the purposes of the present paper it suffices to treat  
the double differential distribution in $B_R,B_L$ as the product of
two independent jets,
\begin{equation}
\label{eq:fact}
   \sigma^{-1}\frac{d^2\sigma}{dB_RdB_L} 
\>=\>  \sigma^{-1}\frac{d\sigma}{dB_R} \>
\sigma^{-1}\frac{d\sigma}{dB_L}
= \Sigma'_1(B_R)\cdot \Sigma'_1(B_L) \>.
\end{equation}
In this approximation the total and the wide broadening distributions
are related to $\Sigma_1$ through
\begin{eqnarray}
  \label{eq:Tdef}
\frac{d\Sigma_T(B)}{dB} &=& \int_0^B
dB_1\>\Sigma'_1(B_1)\>\Sigma'_1(B-B_1)\>, \\
  \label{eq:Wdef}
\frac{d\Sigma_W(B)}{dB} &=& 2\cdot \Sigma_1'(B)\int_0^B dB_N\,
\Sigma'_1(B_N)\>; \qquad 
\Sigma_W(B) = \Sigma^2_1(B)\>.
\end{eqnarray}

\paragraph{Single-jet distribution.}
The Mellin image of the single-jet $B$ distribution, $\sigma(\nu)$, is
defined as
\begin{equation}
\label{Mellin}
\begin{split}
\Sigma'_1(B) 
\>&=\> \int\frac{d\nu}{\pi i}\> e^{2B\nu}\> \sigma(\nu) \>, \\
\Sigma_1(B) 
\>&=\> \int\frac{d\nu}{2\pi i\,\nu}\> e^{2B\nu}\> \sigma(\nu) \>, 
\end{split}
\end{equation}
where the contour runs parallel to the imaginary axis
(in the second case we place the contour at $\Re\nu>0$ in order to
satisfy the normalisation condition \eqref{eq:Snorm}, see below). 
$\sigma(\nu)$ is a regular function of $\nu$ in the entire complex
plane. It is limited (decreases) at $\Re\nu\to+\infty$ which
results in 
$$
 \Sigma_1'(B<0)=\Sigma_1(B<0) \>=\> 0\>. 
$$
In the left half-plane it increases exponentially,
\begin{equation}
  \sigma(\nu) \>\propto\> e^{-2B_m\nu}\>,  \qquad \Re\nu\to \>-\infty\>,
\end{equation}
with $B_m$ the maximal value of single-jet broadening. 
This ensures the kinematical constraints,
\begin{eqnarray}
  \label{eq:Snorm}
 \Sigma_1(B\ge B_m)&=& \sigma(0) \>\equiv\> 1\>,\\
  \label{eq:kincon}
\Sigma_1'(B\ge B_m) &=& 0\>.
\end{eqnarray}

\paragraph{Total broadening distribution.}
The Mellin representation for the integrated total broadening distribution 
in the factorisation approximation follows from \eqref{eq:Tdef}: 
\begin{equation}
\label{eq:Tdist}
 \Sigma_T(B) = \int_{c-i\infty}^{c+i\infty} 
  \frac{d\nu}{2\pi i\,\nu}\> e^{2B\nu}\> \sigma^2(\nu) \>. 
\end{equation}

\paragraph{Mean single-jet broadening.}
To calculate $\lrang{B}$ we average $B$ with the differential
distribution to write
\begin{equation}
\label{eq:meanB}
  \lrang{B}_1 \equiv \int_0^{B_m} BdB\> \Sigma'_1(B)\>.
\end{equation}
The property \eqref{eq:kincon} allows us to extend the $B$ integration
in \eqref{eq:meanB} to infinity, provided that the $\nu$-contour has
been shifted to run to the {\em left}\/ of the imaginary axis,
$\Re\nu<0$:
\begin{equation}
\label{eq:B1mean}
  \lrang{B}_1 =  \int_0^{\infty} BdB\> 
\int_{-c-i\infty}^{-c+i\infty} \frac{d\nu}{\pi i}\> e^{2B\nu}\> \sigma(\nu) 
= \int_{-c-i\infty}^{-c+i\infty} 
\frac{d\nu}{4\pi i}\> \frac{\sigma(\nu)}{\nu^2} 
= -\half \> \sigma'(0)\>.
\end{equation}

\paragraph{Mean total broadening.}
The representation \eqref{eq:Tdist} immediately leads to
\begin{equation}
\label{eq:BTmean}
 \lrang{B}_T \>=\> -\half \left. \frac{d}{d\nu} \sigma^2(\nu)\right|_{\nu=0}
= -\sigma'(0)
= 2 \lrang{B}_1 \>.    
\end{equation}

\paragraph{Mean wide-jet broadening.}
The mean $B_W$, according to \eqref{eq:Wdef}, is given by the integral
\begin{equation}
\label{eq:Wmean}
\begin{split}
\lrang{B}_W&=  \int_0^{B_m} BdB\> \frac{d\Sigma^2_1(B)}{dB}
=-\int_0^{B_m} dB\left\{2[\Sigma_1(B)-1]+[\Sigma_1(B)-1]^2\right\}\\
&=  \int\limits_{\Re\nu<0}\frac{d\nu}{2\pi i}\>\frac{\sigma(\nu)}{\nu^2}
\left\{ 1+ \frac12\int\limits_{\Re\nu_1<0}\frac{d\nu_1}{2\pi i} 
\frac{\nu\>\sigma(\nu_1)}{\nu_1(\nu+\nu_1)}\right\} \\
&= \int\frac{d\nu}{2\pi i}\>
\frac{\sigma(\nu)\left[\,1+\sigma(-\nu)\,\right]}{2\,\nu^2}
\>=\>  - \half\sigma'(0) +\int_{-i\infty}^{i\infty} 
  \frac{d\nu}{2\pi i} \>\frac{\sigma(\nu)\sigma(-\nu)}{2\,\nu^2} .
\end{split}
\end{equation}
The latter integral cannot be evaluated by residues, since the
integrand exponentially increases in both directions,
$\sigma(\nu)\sigma(-\nu)\propto\exp(|\Re\nu| B_{m})$.

\subsection{Kinematics of small $B$}
Here we introduce the kinematical variables needed to analyse 
$B$ in the soft limit.
We use the Sudakov representation for the momenta of the primary 
quark and  antiquark $p_1$ and $p_2$ and of the accompanying soft 
partons $k_i$
\begin{equation}
  \label{Sud}
\begin{split}
  p_1 &= A_1 P+B_1 \bP+p_{t1}\>, \\
  p_2 &= A_2 P+B_2 \bP +p_{t2} \>, \\
  k_i &= \al_{i}P + \be_{i}\bP + k_{ti}\>.
\end{split}
\end{equation}
The two light-like Sudakov vectors $P$ and $\bP$ are taken along 
the thrust axis and their sum is the total incoming momentum ($2P\bP=Q^2$).

In the soft region all quantities 
\[
\al_i,\>\be_i,\>(1-A_1),\>(1-B_2)\>,\frac{k_{ti}}{Q}\>
\]
are small and of the same order while $A_2$ and $B_1$ are much smaller 
(quadratic in transverse momentum)
and will be neglected.
As shown in Appendix~\ref{Ap:PNPrad}, for the two-loop {\em perturbative}\/ 
contribution one can approximate the quark-antiquark direction with 
that of the thrust axis, i.e. one can neglect $p_{t1}$ and $p_{t2}$.  
However, as explained above, the {\em non-perturbative}\/ contribution is 
sensitive to the difference between the thrust and quark axis and 
then we shall take into account $p_{ti}$.
We introduce the quark and antiquark angular variables 
(rescaled quark transverse momenta)
\begin{equation}
  \label{eq:pdef}
  \vec{p} \>\equiv\> \frac1{A_1}\, \vec{p}_{t1}\>,
\qquad
  \vec{\bar p} \>\equiv\> \frac1{B_2}\, \vec{p}_{t2}\>.  
\end{equation}
The soft radiation matrix element depends on the invariant
\[
\frac{2(p_1k_i)(p_2k_i)}{(p_1p_2)}\>\simeq\>
\frac{1}{k^2_{ti}}\>
(\vec{k}_{ti}-\al_i \vec{p})^2
(\vec{k}_{ti}-\be_i \vec{\bar p})^2\>,
\]
where the first (second) factor is the squared transverse momentum 
of $k_i$ with respect to the quark (antiquark) direction.

In the following we concentrate on soft partons emitted within
the right hemisphere, near the quark direction. 
The kinematical constraint for such a parton is $\al_i>\be_i$. 
Since a massless emitted parton satisfies $\al_i\be_i=k^2_{ti}/Q^2$, 
\begin{equation}
  \label{eq:rhem}
\al_i>k_{ti}/Q\>.  
\end{equation}
For such a right-hemisphere parton one can neglect the difference between 
the thrust and antiquark directions 
$(\vec{k}_{ti}-\be_i\vec{\bar p}\simeq \vec{k}_{ti})$, so that 
the matrix element depends only on the transverse momentum with respect 
to the quark axis
\begin{equation}
  \label{tktdef}
 \vtkti{i}\>\equiv\>\vec{k}_{ti}-\al_i \vec{p}\>.
\end{equation}
According to the definition of the thrust axis, the vector sum of the 
parton transverse momenta in the right (or left) hemisphere is zero and
we can write
\begin{equation}
  \label{eq:4}
 0= \vec{p}_{t1}+\sum_{i\in R} \vec{k}_{ti}=  
 A_1\vec{p} + \sum_{i\in R} \vec{k}_{ti}
\simeq  \vec{p} + \sum_{i\in R} \vtkti{i}\>.  
\end{equation}
where we have neglected  the small $\al$-components of the 
{\em left}-jet partons ($1-A_1\simeq \sum_{\in R}\al_i$).

The right-jet broadening $B_R$ is defined by
\begin{equation}
  \label{BRdef}
  2B_R Q\>=\> p_{t1} + \sum_{i\in R} k_{ti} 
\>=\>  A_1 p +\sum_{i\in R} k_{ti}
\>\simeq\> p +\sum_{i\in R}(k_{ti}-\al_i p)\>,
\end{equation}
where again we have neglected  the small $\al$-components of the 
{\em left}-jet partons.

Similarly one introduces $B_L$, the left broadening,
\begin{equation}
  \label{BLdef}
  2B_L Q \>\simeq\> \bar{p} +\sum_{i\in L}(k_{ti}-\be_i \bar{p})\>.
\end{equation}

\subsection{Resummation and the radiator}

Here we recall the resummed expression for the broadening 
distribution at small $B$~\cite{DLMSbroad}. 
In this region the distribution is obtained 
by resumming contributions 
from the emission of any number of soft partons and can be expressed
in terms of a ``radiator''. 
The two-loop analysis of the radiator is presented in detail 
in Appendix~\ref{Ap:PNPrad}. 

The {\em integrated}\/ single-jet broadening distribution (e.g. the
right jet) is given by
\begin{equation}
  \label{SigR}
\Sigma_1(B)= Q^2\>\sum_n\int \frac{d\sigma_n}{\sigma} \cdot 
\delta^2(\vec{p} + \sum_{i\in R} \vtkti{i})\>
\Theta\left( 2BQ - p - \sum_{i\in R}(k_{ti}- \al_i p)
\right),
\end{equation}
where we have used \eqref{eq:4} to represent the transverse momentum
constraint in terms of the rescaled quark momentum $\vec{p}$
\eqref{eq:pdef} and the secondary parton transverse momenta {\em with
respect to the quark direction}, $\vtkti{i}$.
The factors  
$d\sigma_n/\sigma$ are the $n$-parton emission distributions which 
factorise for $B\ll1$, i.e. in the soft limit. 
To perform the sum over $n$ one needs to factorise also the constraints.
This is obtained by using the Fourier and Mellin representations for the 
delta and theta functions 
\begin{multline}
\label{uki}
\delta^2(\vec{p} + \sum_{i\in R} \vtkti{i})\>
\Theta\left(2BQ-p-\sum_{i\in R}(k_{ti}-\al_i p)\right)
\\
=\int \frac{d^2b}{(2\pi)^2}\>e^{-i\vec{b}\vec{p}}
\int \frac{d\nu}{2\pi i \nu}\> e^{2B \nu}\> e^{-\nu p/Q}\>\prod_{i\in R}\>
e^{-\nu(k_{ti}-\al_ip)/Q}e^{-i\vec{b}\vtkti{i}}\>.
\end{multline}
where we have introduced the vector impact parameter $\vec{b}$ 
conjugate to the transverse momentum with respect to the quark axis
\eqref{tktdef}. One can now resum soft-parton emissions in the Mellin
representation \eqref{Mellin}. The resulting integrated single-jet
broadening distribution is
\begin{equation}
\label{Sigma}
\Sigma_1(B)= \int \frac{d\nu}{2\pi i\, \nu} e^{2\nu B} \>\sigma(\nu) \>, 
\qquad
\sigma(\nu) = \int \frac{d^2p\, d^2b}{(2\pi)^2}\> 
e^{-i\vec{b}\vec{p}}\>e^{-\nu p/Q}\>e^{-\cRt(\nu,b;p)}\>,
\end{equation}
The exponent $\cRt$ in \eqref{Sigma} is the ``radiator'' analysed 
in Appendix~\ref{Ap:PNPrad}. 
Since soft real and virtual partons are responsible also for the 
non-perturbative effects, the radiator contains, together with the 
\PT\ 
contribution, also a \NP\ 
correction,
\begin{equation}
  \label{eq:cRfull}
  \cRt(\nu,b;p) =   \cR^{(\PT)}  +  \cR^{(\NP)}\>.
\end{equation}

\paragraph{PT component.}
At two-loop accuracy, the \PT\ part of the radiator is
$p$-independent.  At large values of $\nu$ and/or $bQ$, it is a
function of the combination~\cite{DLMSbroad}
\begin{equation}
  \label{eq:mudef}
\bmu = e^{\gamma_E}\,\frac{\nu+\sqrt{\nu^2+(bQ)^2}}{2} \>,
\end{equation}
and reads
\begin{equation}
  \label{eq:cRone}
\cR^{(\PT)}=  \cR(\bmu)\>=\> \int_{Q/\bmu}^Q \frac{dk}{k}\> \cR'(Q/k)\>, \quad
  \cR'(\bmu)\>=\>\frac{2C_F}{\pi}\as(Q/\bmu)\left(\ln{\bmu}-\frac34\right).
\end{equation}
The running coupling is taken in the physical scheme~\cite{CMW}.  This
expression accommodates all terms $\as^n\ell^{n+1}$ and $\as^n\ell^n$
with $\ell=\ln \bmu$.  We present $\cR$ as a function of a single
dimensionless parameter $\bmu$ but it is implied that it also contains
a $Q$-dependence via the running coupling $\as(k_t)$, with
$Q/\bmu<k_t<Q$.

\paragraph{NP component.}
According to the procedure developed in~\cite{DMW}, the leading
power-suppressed \NP\ contribution to the radiator has the following general
form, see Appendix~\ref{Ap:PNPradN} for details,
\begin{equation}
  \cR^{(\NP)} =\cM \frac{C_F}{\pi} \int_0^\infty \frac{dm^2}{m^2}\>
\delta\ae(m^2) \cdot \Omega_0(m^2)\>.
\end{equation}
Here $\cM$ is the Milan factor which emerges from the two-loop
analysis, and
$\delta\ae(m^2)$ is the \NP\ 
component of the {\em effective coupling}\/ related to the standard
$\as$ by the dispersive relation, with $m^2$ the corresponding
dispersive variable acting as the gluer's ``mass''.

The factor $\Omega_0(m^2)$ is a ``trigger function'' which is specific
to a given observable.  A power-behaved \NP\ %
contribution is determined by the leading non-analytic in $m^2\!\to\!0$
term in $\Omega_0$.  For the broadening measure, as well as for other
jet shapes, the leading non-analytic piece, $\delta\Omega_0$, is
proportional to $\sqrt{m^2}$ and is given by the following expression,
\begin{equation}
  \label{eq:111}
\delta\Omega_0(m^2) = \nu\frac{\sqrt{m^2}}{Q}\cdot
\int_{\tpt/Q}^{{\tpt}/{\sqrt{m^2}}}
\frac{du}{u}\>  \int_0^{2\pi}\frac{d\psi}{2\pi}
\left(\sqrt{1+u^2+2u\cos\psi}-u\right) \>.
\end{equation}
Here 
$$
u=\frac{\al\tpt}{\sqrt{m^2}} = \frac{\tpt}{Q}e^\eta \>, \qquad
0\le\eta < \ln\frac{Q}{m}\>, 
$$ 
with $\eta$ the gluer rapidity with respect to the thrust axis. 
The expression in the brackets in \eqref{eq:111}
accounts for the mismatch between the quark and thrust axes.  
Since, with account of the  azimuthal integration, the integrand 
falls rapidly at $u\to\infty$, the rapidity-integral, in the $m=0$
limit, has a finite value proportional to $\ln p$.
Thus the \NP\ %
component of the radiator becomes
\begin{equation}
   \cR^{(\NP)} = \nu\cdot \cP\ln\frac{p_0}{p}\>,\quad 
\eta_0 \equiv \ln\frac{p_0}{Q}   \simeq -0.6137056\>.
\end{equation}
The definition of the \NP-parameter $\cP\propto 1/Q$ 
introduced in~\cite{DLMSuniv}
is recalled in Appendices~\ref{Ap:PNPradN} and~\ref{sec:Collection}.

We notice that if in \eqref{eq:111} we formally set  
$\tpt\equiv0$, which corresponds 
to disregarding the perturbative quark recoil,
we reconstruct the old {\em wrong}\/ answer~\cite{DLMSuniv} 
\begin{equation}
\label{eq:old2}
\left . \delta\Omega_0(m^2) \right|_{p\equiv 0}
\>\simeq\> \nu\frac{\sqrt{m^2}}{Q} \int_0^{\ln Q/m} d\eta
\> \Longrightarrow\>
\nu\frac{\sqrt{m^2}}{Q} \cdot
\left(\ln\frac{Q}{\sqrt{m^2}}-\frac34\right),  
\end{equation}
where we have restored the $3/4$ ``hard correction'' 
coming from the region $\al\sim1$.

In the following we recall the perturbative part of the
$B$ distribution, while the non-perturbative one will be treated in
the next section. 

\subsection{Recollection of the perturbative result}
The $p$-integration giving the perturbative part of the single-jet
distribution $\sigma(\nu)$ in \eqref{Sigma} is explicitly carried out
in Appendix~\ref{Ap:PNPradP}, see \eqref{eq:Sigpty},
\begin{equation}
  \label{eq:Sig}
\sigma^{(\PT)}(\nu) = \int \frac{d^2p\, d^2b}{(2\pi)^2}\> 
e^{-i\vec{b}\vec{p}}\>e^{-\nu p/Q}\>e^{-\cR(\bmu)}
=  \int_1^\infty \frac{dy}{y^2}\>e^{-\cR(\bmu)} , \quad
   \bmu=e^{\gamma_E}\nu\frac{1+y}{2} \>.
\end{equation}
To evaluate the Mellin integral we introduce the following convenient
operator representation which we shall use extensively below,  
\begin{equation}
\label{eq:COR}
  e^{-\cR(\bmu)} = \left. e^{-\cR(e^{-\partial_a})}\>
    (\bmu)^{-a}\right|_{a=0}. 
\end{equation}
The Mellin image of the integrated single-jet distribution can then be 
represented as
\begin{equation}
  \label{eq:sigPTnuop}
  \sigma^{(\PT)}(\nu) = \left. e^{-\cR(e^{-\partial_a})}\>
\left(e^{\gamma_E}\lam(a)\right)^{-a}\cdot \nu^{-a}\right|_{a=0},
\end{equation}
where we have introduced the function \cite{DLMSbroad}
\begin{equation}
\label{eq:lamdef}
 \lam^{-a}\>\equiv\>
 \int_1^\infty \frac{dy}{y^2}\>\left(\frac{1+y}{2}\right)^{-a}
= \int_0^1 dz \left(\frac{1+z}{2z}\right)^{-a} ; \qquad
 \lam(0)=2\>, \quad \lam(\infty)=1\>.
\end{equation}
Performing the $\nu$-integration we arrive at
\begin{equation}
\label{eq:Sigop}
 \Sigma^{(\PT)}_1(B) =  \int\frac{d\nu}{2i\pi\nu}\>e^{2B\nu}\> 
\sigma^{(\PT)}(\nu) = \left. e^{-\cR(e^{-\partial_a})}\>
\frac{1}{\Gamma(1+a)}\>\left(\frac{e^{\gamma_E}\lam(a)}{2B}\right)^{-a} 
\right|_{a=0}.
\end{equation}
Using the identity
$$
   \left. e^{-\cR(e^{-\partial_a})}\> x^{-a}g(a) \right|_{a=0} \>=\> 
   \left.  e^{-\cR(xe^{-\partial_a})} \>g(a)\right|_{a=0} \>,
$$
we can absorb the power factor $x^{-a}$,
\begin{equation}
  \label{eq:xdef}
 x=\frac{\lambda(\cR') e^{\gamma_E}}{2B}\>, \quad
   \cR'=\cR'(x)=\frac{d\cR(x)}{d\ln x} \>,
\end{equation}
into rescaling of the argument of the perturbative radiator:
\begin{equation}
   \Sigma^{(\PT)}_1(B) \>=\> e^{-\cR(xe^{-\partial_a})}\>
\left. 
\frac{1}{\Gamma(1+a)}\>\left(\frac{\lam(\cR')}{\lambda(a)}\right)^{a}
\right|_{a=0}\>.
\end{equation}
Performing the logarithmic expansion of the radiator,  
$$
 -\cR\left(xe^{-\partial_a}\right) \>=\> -\cR(x) + \cR'(x)\partial_a
  -\half\cR''(x)\partial_a^2+ \ldots \>, 
$$
we conclude that the action of the operator on a function which is
{\em regular}\/ in the origin reduces to substituting $\cR'(x)$ for
$a$, while $\cR''(x)=\cO{\as}$ and higher derivatives produce
negligible corrections:
\begin{equation}
  \label{eq:reg}
\left.  e^{-\cR(xe^{-\partial_a})} \>g(a)\right|_{a=0}
\>=\> e^{-\cR(x)}\,g(\cR'(x))\>\left(1+\cO{\cR''(x)}\right).
\end{equation}
Thus we can evaluate \eqref{eq:Sigop} as
\begin{equation}
\label{eq:SigmaPTfin'}
\Sigma^{(\PT)}_1(B)\>=\> \frac{e^{-\cR(x)}}{\Gamma(1+\cR'(x))}
\left(1+\cO{\as}\right),
\end{equation}
where the parameter $x$ is a function of $B$ implicitly defined by 
\eqref{eq:xdef}. 
This expression can be simplified. To this end we observe that 
within our accuracy we can substitute $B^{-1}$ for $x$ in $\cR'$,
$$
 \cR'(x) = \cR'(B^{-1}) + \cO{\as^{n+1}\ln^n B}\>, \quad n\ge 0\,; 
\qquad  \cR'(B^{-1}) \simeq \frac{2C_F}{\pi}\as(QB)\ln B^{-1} \,.
$$
At the same time, the finite product $xB$ should be kept in the
exponent $\cR(x)$.
We finally obtain
\begin{equation}
\label{eq:SigmaPTfin}
\Sigma^{(\PT)}_1(B)\>=\> 
\frac{e^{-\cR(\bB ^{-1})}}{\Gamma(1+\cR')}\>,
\qquad \bB= \frac{2B}{e^{\gamma_E}\lambda(\cR')}\>,
\qquad \cR'=\cR'(B^{-1})\>.
\end{equation}
For the wide-jet and total broadening distributions we obtain, in a
similar way,
\begin{equation}
  \label{eq:SigmasPT}
  \Sigma_W(B) = \frac{e^{-2\cR(\bB^{-1})}}{\Gamma^2(1+\cR')} \>, \qquad
  \Sigma_T(B) = \frac{e^{-2\cR(\bB^{-1})}}{\Gamma(1+2\cR')} \>.
\end{equation}
These results are equivalent to those based on the steepest descent
evaluation presented in~\cite{DLMSbroad}.
The resummed \PT\ %
answers can be given in various forms which are equivalent at the
level of the leading $\as^n\ln^{n+1} B$ and next-to-leading
$\as^n\ln^{n} B$ contributions to the exponent, see, e.g. equations
(4.25) and (4.26) of \cite{DLMSbroad}.  The specific form which should
be used for matching with the exact second order \PT\ %
answer will be discussed in Appendix~\ref{sec:Collection}.

\section{Non-perturbative correction to the $B$ distribution}

With account of the leading \NP\ %
contribution the integrated single-jet $B$ distribution takes the form
(cf.\ \eqref{eq:Sig})
\begin{equation}
  \label{Sig}
\sigma(\nu) \>=\>  \int_0^\infty bdb\>e^{-\cR(\bmu)}\>
  \int_0^\infty pdp\>e^{-\nu \tpt/Q}
  \left(\frac{\tpt}{p_0}\right)^{\nu\eps} J_0(b\tpt)\,.
\end{equation}
Performing the $p$-integration, see \eqref{eq:nptint} 
in Appendix~\ref{Ap:PNPradN},  
the distribution assumes the form
\begin{equation}
\begin{split}
  \label{eq:SigMel}
  \sigma(\nu) \>&=\>  \sigma^{(\PT)}(\nu)\>+\> \nu\eps\, f(\nu) \>+\> 
\cO{\eps^2}\>,
\end{split}
\end{equation}
with the non-perturbative correction given by the following expression
\begin{equation}
\begin{split}
  \label{eq:fdef}
f(\nu) 
&=\> \int_1^\infty \frac{dy}{y^2} \>e^{-\cR(\bmu)}
  \left(2-\gam_E
-\eta_0
+\ln\frac{1+y}{2
\nu y^2} -y\right) ; \qquad y\equiv \sqrt{1+\left(\frac{bQ}{\nu}\right)^2}. 
\end{split}
\end{equation}
We recall that
\begin{equation}
  \bmu = \half e^{\gamma_E}\left(\nu+\sqrt{\nu^2+(bQ)^2}\right) \>=\> 
\nu\, e^{\gamma_E} \frac{1+y}{2}\>.
\end{equation}
The function $f(\nu)$ will determine the \NP\ %
corrections to the $B$ distributions and to the means. In particular,
the \NP\ %
effects in the distributions are given by the inverse Mellin transform
of $f(\nu)$ which we calculate in terms of the operator technique
introduced in \eqref{eq:COR}.  In the case of the \PT\ %
distribution, the operator $\exp\{-\cR(xe^{-\partial_a})\}$ acted upon
a function $g(a)$ {\em regular}\/ at $a\!=\!0$, see \eqref{eq:reg}.
In the case of $B_T$, as we shall see later, we need to calculate the
operator $\exp\{-\cR(e^{-\partial_a})\}$ acting upon a function
$g(a)\propto 1/a$ which is {\em singular}\/ in the origin.  To this
end we introduce and evaluate in Appendix~\ref{Ap:E(x)} the function
\begin{equation}
  \label{eq:Edef}
  E(x) \>=\> \left.  e^{\cR(x)}\,
e^{-\cR(xe^{-\partial_a})} \> a^{-1}\right|_{a=0} 
\>=\> \int_x^\infty \frac{dz}{z}\> e^{\cR(x)-\cR(z)} \>.
\end{equation}
The value $E(1)$ enters into the expression for the power correction
to $\lrang{B}$, 
\begin{equation}
  \label{eq:E1}
E(1) = \frac{\pi}{2\sqrt{C_F\as}} +\frac{3}{4} -\frac{\beta_0}{6C_F}
\>+\> \cO{\sqrt{\as}}\>.
\end{equation}
As far as the \PT\ %
component of the distribution $\sigma^{(\PT)}(\nu)$ in
\eqref{eq:SigMel} is concerned, we have derived the resummed
expression which is valid for large values of $\nu$ and determines the
$B$ distributions in the $B\ll1$ two-jet kinematics (soft limit).  The
$B$ distributions at large $B$, as well as the means $\lrang{B}$, are
determined by finite moments $\nu$. The soft resummation programme is
irrelevant here, and the exact fixed-order analyses should be carried
out instead.
At the same time, the \NP\ %
component $f(\nu)$ in \eqref{eq:SigMel} which originates from {\em
  soft gluers}\/ remains applicable both for $B\ll1$ distributions and
the means, that is for arbitrary values of $\nu$.
Having said that we turn to the calculation of the leading \NP\ 
effects in mean broadenings.

\subsection{Power corrections to means}
Substituting the sum of perturbative and non-perturbative components,
\eqref{eq:SigMel}, into the expressions for the mean broadening 
\eqref{eq:B1mean} and \eqref{eq:BTmean} we obtain
\begin{equation}
\label{eq:mean1Tnp}
 \lrang{B}_T -  \lrang{B}_T^{(\PT)} =
2\left(  \lrang{B}_1 -  \lrang{B}_1^{(\PT)} \right) 
\>=\> -\eps\, f(0)\>.
\end{equation}
The value of $f$ in the origin is related with the function $E$ 
\eqref{eq:Edef} in Appendix~\ref{Ap:f(0)},
\begin{equation}
  f(0)\>=\> -\left( E(1) \>+\> \eta_0\right) 
\left[\, 1 \>+\> \cO{\as}\,\right] .
\end{equation}
This results in 
\begin{equation}
\begin{split}
\label{eq:npmean1}
\lrang{B}_1 - \lrang{B}_1^{(\PT)} 
\>\simeq\> \frac{\eps}{2}
\left( \frac{\pi}{2\sqrt{C_F\as}} +\frac{3}{4} -\frac{\beta_0}{6C_F}
+ \eta_0\right).
\end{split}
\end{equation}

\paragraph{Mean wide-jet broadening}
 is given in terms of the Mellin integral in \eqref{eq:Wmean}. 
Extracting the non-perturbative component we obtain
\begin{equation}
\label{eq:npcompW}
\begin{split}
\lrang{B}_W &=
\lrang{B}^{(\PT)}_W  +\eps\left[\,\delta-\half f(0)\,\right] \>+\>
\cO{\eps^2}, 
\end{split}
\end{equation}
with
\begin{equation}
 \delta \>=\>  \frac12 \int_{-i\infty}^{i\infty}   \frac{d\nu}{2\pi i\nu} 
\left[\, f(\nu)\sigma^{(\PT)}(-\nu)- f(-\nu)\sigma^{(\PT)}(\nu)\,\right].
\end{equation}
This integral is evaluated in Appendix~\ref{Ap:delta} resulting in 
\begin{equation}
\delta \>=\>  \left. \half\left(e^{-2\cdot\cR(e^{-\partial_a})} 
- e^{-\cR(e^{-\partial_a})}\right) a^{-1} \right|_{a=0}.  
\end{equation}
We note that the first operator here
corresponds to $E(1)$ evaluated with $2\cR$ substituted for the radiator. 
This leads to
\begin{equation}
\begin{split}
\label{eq:meanWIDE}
 \lrang{B}_W - \lrang{B}_W^{(\PT)} &=\>  \frac{\eps}{2} \left( \left. 
e^{-2\cdot\cR(e^{-\partial_a})}  a^{-1} \right|_{a=0}
+\eta_0 \right)\left[\, 1 \>+\> \cO{\as}\,\right] \\
\>&\simeq\>  \frac{\eps}{2} \left( \frac{\pi}{2\sqrt{2C_F\as}} 
 +\frac{3}{4} -\frac{\beta_0}{12C_F} + \eta_0\right).
\end{split}
\end{equation}
The result can be obtained from that for the single-jet (total) broadening 
by the simple substitution $C_F\to 2C_F$ in \eqref{eq:npmean1}. 

\subsection{NP shift in the single- and wide-jet distributions} 
Having obtained the $1/Q$ power corrections to the means we now turn
to the analysis of the \NP\ effects in the distributions in the
$B\ll1$ region.  To this end we need to perform the inverse Mellin
transform of the distribution \eqref{eq:SigMel} containing the \NP\ %
term $f(\nu)$ given in \eqref{eq:fdef}.

To perform the integration over $y$ (related to the impact parameter 
$b$, see \eqref{eq:fdef}) we make use of the operator representation 
\eqref{eq:COR} to write
$$ 
 f(\nu) =  \left. e^{-\cR(e^{-\partial_a})}\> 
\int_1^\infty \frac{dy}{y^2} 
\left(e^{\gamma_E} \frac{1+y}{2}\right)^{-a}  
\left(2-\gam_E-\eta_0 +\ln\frac{1+y}{2y^2} +\partial_a -y\right)
\cdot \nu^{-a}  \right|_{a=0}.
$$ 
The result of the $y$-integration can be represented as follows:
$$ 
 f(\nu)= \left. 
e^{-\cR(e^{-\partial_a})}\> \left(e^{\gamma_E}\lambda\right)^{-a}   
\left(2-\gam_E-\ln 2-\eta_0 +\rho(a) 
- \chi(a) +\partial_a -\frac1a \right)
\nu^{-a}  \right|_{a=0},
$$ 
where $\lambda=\lambda(a)$ is defined in \eqref{eq:lamdef} and 
 $\rho$ and $\chi$ are given by the integrals
\begin{equation}
\begin{split}
  \rho(a) &= \int_1^\infty \frac{dy}{y^2} 
\left(\frac{1+y}{2\lambda}\right)^{-a}  \ln\frac{1+y}{y^2} \>, \\
  \chi(a) &=  - \frac{1}a + \int_1^\infty \frac{dy}{y} 
\left(\frac{1+y}{2\lambda}\right)^{-a}\>=\> 2\frac{\lam^a-1}{a}\>. 
\end{split}
\end{equation}
The functions  $\rho(a)$ and $\chi(a)$ are regular at $a=0$ and  
vary slowly between $\rho(0)=-2+2\ln2$, $\rho(\infty)=\ln2 $ 
and $\chi(0)=2\ln2 $, $\chi(\infty)=2$ 
correspondingly.
Introducing the function
$$
 C(a) \>=\> 2-\gam_E-\ln 2 -\eta_0 +\rho(a) - \chi(a) \>, 
$$
the operator expression for the \NP\ %
component of the distribution takes the form
\begin{equation}
  \label{eq:fnuoper}
 f(\nu)= \left. 
e^{-\cR(e^{-\partial_a})}\> \left(e^{\gamma_E}\lambda\right)^{-a}   
\left(C(a) +\partial_a -\frac1a \right) 
\nu^{-a} \right|_{a=0}\>.
\end{equation}
The $\nu$-integration of \eqref{eq:SigMel} is now readily
performed. For the single-jet distribution we have
\begin{equation}
\label{eq:Rop}
\begin{split}  
\Sigma_1(B)&= \left. 
e^{-\cR(e^{-\partial_a})}
{( e^{\gamma_E}\lambda)^{-a}} 
\left[\, \frac1{\Gamma(1+a)} + \frac{\eps}{2B}
\left(C(a) +\partial_a-\frac1a\right)\frac1{\Gamma(a)}\right] 
\>{(2B)^a} \right|_{a=0}.
\end{split}
\end{equation}
We observe that the potential singularity at $a=0$ cancels in the combination 
$$
\left(C(a)+\partial_a-\frac1a\right)\frac{(2B)^a}{\Gamma(a)} 
= 
-a\frac{(2B)^a}{\Gamma(1+a)}\cdot \tC(B,a)\>,
$$
with
$$ 
  -\tC(B,a) \>=\> C(a) +\ln 2B - \psi(1+a) \>=\> \ln B -\eta_0
+2 + \rho(a)-\chi(a)+ \psi(1)\!-\!\psi(1+a)
$$
a function regular in the origin, $a\!=\!0$. The function $\psi(x)$
is defined as $\psi(x) = d\ln \Gamma(x)/dx$.  We finally arrive at
\begin{equation}
\label{eq:singjetshift}
\begin{split}  
\Sigma_1(B) &= \left. e^{-\cR(e^{-\partial_a})}
\frac{\left(\frac{ e^{\gamma_E}\lambda(a)}{2B}\right)^{{-a}}} {\Gamma(1+a)} 
\left[1-\frac{a\,\eps}{2B} \tC(B,a) \,\right]  \right|_{a=0} \\
&= \left. e^{-\cR\left(xe^{-\partial_a}\right)}\>
\frac{ \left(\frac{\lambda(\cR')}{\lambda(a)}\right)^{a}}{\Gamma(1+a)}
\left[\, 1-\frac{a\,\eps}{2B} \tC(B,a)\,\right] \right|_{a=0} .
\end{split}
\end{equation}
Here we have absorbed the power factor $x^{-a}$ with $x$ defined in
\eqref{eq:xdef} into a rescaling of the argument of the \PT\ %
radiator as we did before.  As in the case of the \PT\ %
distribution considered above, the action of the operator on a {\em
  regular}\/ function results in substituting
$\cR'(x)\simeq\cR'(B^{-1})$ for $a$, according to \eqref{eq:reg}.  We
get
\begin{equation}
\Sigma_1(B) 
= \frac{e^{-\cR(\bB^{-1})}}{\Gamma(1+\cR')}\cdot
\left[\, 1-\cR'\cdot \frac{\eps}{2B}\,\tC(B,\cR') \,\right] , 
\end{equation}
with $\bB$ defined in \eqref{eq:SigmaPTfin}.
This correction can be cast as a $B$-dependent $1/Q$ {\em shift}\/
of the perturbative distribution, namely
\begin{equation}
\Sigma_1(B) \>\simeq\> \Sigma^{(\PT)}_1\left(B - \frac{\eps}{2}D_1(B)\right), 
\end{equation}
with 
\begin{equation}
  \label{eq:D1ans}
D_1(B) = \tC(B,\cR') \>=\> \ln B^{-1} +\eta_0 -2 - \rho(\cR')+\chi(\cR')
+ \psi(1+\cR')\!-\!\psi(1)\,,\quad  \cR'=\cR'(B^{-1})\>.
\end{equation}
For not too small values of $B$ such that $\cR'\ll 1$ the
shift $D_1$ assumes a simple form
\begin{equation}
  \label{eq:D1small}
 D_1 \>\simeq\> \ln B^{-1} + \eta_0\>, \quad  \cR'\ll1\>,
\end{equation}
while in the opposite limit of extremely small $B$,
\begin{equation}
  \label{eq:D1large}
  D_1 \>\simeq\> \ln\frac{\cR'}{2B} +\gamma_E + \eta_0\>,  \quad \cR'\gg1\>.
\end{equation}
The same shift applies also to the {\em wide}-jet broadening 
distribution which, to the needed accuracy, is simply given by the squared
single-jet distribution according to \eqref{eq:Wdef}.

\subsection{
  NP shift in the total broadening distribution} To obtain the
integrated distribution for the total broadening, $\Sigma_T(B)$, we
need to perform the inverse Mellin transform of $\sigma^2(\nu)$ in
\eqref{eq:Tdist}.  Invoking the operator representations
\eqref{eq:sigPTnuop} and \eqref{eq:fnuoper} for $\sigma^{(\PT)}(\nu)$
and $f(\nu)$ correspondingly, we construct the product ($a,b\to0$)
\begin{multline*}  
\sigma^2(\nu) \>=\> e^{-\cR(e^{-\partial_a})}e^{-\cR(e^{-\partial_b})}
{\left({e^{\gamma_E}\lambda(a) } \right)^{-a}
      \left({ e^{\gamma_E}\lambda(b)} \right)^{-b}} \\
\left. 
\left[\, 1+ 2\cdot\nu \eps
\left(C(a)+ \partial_a-\frac1a\right) 
\,\right]\nu^{-(a+b)} \right|_{a=b=0}\> +\> \cO{\eps^2},
\end{multline*}
where we have made use of the $a\leftrightarrow b$ symmetry.
Evaluating the $\nu$-integral we obtain
\begin{multline*}  
\Sigma_T(B) \>=\> e^{-\cR(e^{-\partial_a})}e^{-\cR(e^{-\partial_b})}
{\left({ e^{\gamma_E}\lambda(a)} \right)^{-a}
      \left({ e^{\gamma_E}\lambda(b)} \right)^{-b}} \\
\left. \left[\frac{{(2B)^{a+b}}}{\Gamma(1+a+b)}
+2\cdot\frac{\eps}{2B}
\left(C(a)+ \partial_a-\frac1a\right) 
\frac{(2B)^{a+b}}{\Gamma(a+b)}\right]\right|_{a=b=0},
\end{multline*}
Here the first term gives $\Sigma^{(\PT)}(B)$ and the second one
accounts for the leading $1/Q$ correction contribution. 
Recalling the definition of $x$ \eqref{eq:xdef} we write
($\cR'=\cR'(x)$)
\begin{multline}  
  \label{eq:SigTsns}
\Sigma_T(B) \>=\> e^{-\cR(xe^{-\partial_a})}e^{-\cR(xe^{-\partial_b})}
\frac{\left(\frac{\lam(\cR')}{\lam(a)}\right)^{a} 
 \left(\frac{\lam(\cR')}{\lam(b)}\right)^{b}} {\Gamma(1+a+b)} \\
\left. \cdot\left[\,  1+\frac{\eps}{B}\,\Gamma(1+a+b)
\left(C(a)+ \ln2B + \partial_a -\frac1a\right) \frac{1}{\Gamma(a+b)}
\,\right]\right|_{a=b=0} .
\end{multline}
The first factor is identical to that for the perturbative
distribution. The second factor can be represented as
\begin{multline}
\label{eq:STfact}
  1+\frac{\eps}{B}\left[\,(a+b)
\left(C(a)+ \ln2B -\psi(1+a+b) \right)-\frac{b}a\,\right] \\
=  1\>-\>\frac{\eps}{B} \left( (a+b)\left[\, \tC(B,a)
    +\psi(1+a+b)-\psi(1+a)\,\right]+\frac{b}{a}\right).
\end{multline}                                
We observe that, in contrast to the single-jet case
(cf.~\eqref{eq:singjetshift}), we get a correction term {\em
  singular}\/ in $a$, which reminds us of the case of $\lrang{B}$
considered above.  With the unity and the non-singular term in
\eqref{eq:STfact} we proceed as before substituting $\cR'$ for $a$ and
$b$, see~\eqref{eq:reg}.  The factor $a+b$ produces $2\cR'$ which
makes it possible to interpret the correction in terms of a shift.
The contribution of the singular piece in \eqref{eq:STfact} to the
shift is calculated in Appendix~\ref{Ap:DT}. The final result reads
\begin{subequations}
\label{eq:DTans}
\begin{align}
\Sigma_T(B) &= \Sigma_T^{(\PT)}\left(B-\frac{\eps}{2}D_T(B)\right)\>, \\
  \label{eq:DTansb}
D_T(B)  &= 2D_1(B) +2[\psi(1+2\cR')-\psi(1+\cR')] + 
H(\bB^{-1})\>,
\quad \cR'=\cR(B^{-1})\>.
\end{align}
\end{subequations}
Here the single-jet shift $D_1(B)$ is given in \eqref{eq:D1ans}, and
$\bB$ defined in \eqref{eq:SigmaPTfin}.
The \NP\ %
shift in the total broadening distribution, \eqref{eq:DTans}, includes
the function $H(\bB^{-1})$ which is analysed in Appendix~\ref{Ap:DT}.
The shift has a rather complicated $B$ dependence: it changes from
$$
 D_T(B) \>\approx\> \ln \frac1B + \mbox{const}\>, \quad\mbox{for}\>\>\> 
\frac{\as}{\pi}\ln^2B < 1\>,
$$
to
$$
 D_T(B) \>\approx\> 2\cdot\ln\frac1B \>, \quad\qquad\>\mbox{for}\>\>\> 
\frac{\as}{\pi}\ln^2B \gg 1\>.
$$
Indeed, for moderately small values of $B$ such that $\as\ln^2B\!<\!1$
($\cR'\!<\!\sqrt{\as}$), we have (see~\eqref{eq:Hxsmall})
$$
 H(\bB^{-1}) \>\simeq\>  \ln B +  \frac{\pi}{2\sqrt{C_F\as}} 
+\frac{3}{4} -\frac{\beta_0}{6C_F}\>.
$$
Together with \eqref{eq:D1small} this results in
\begin{equation}
  D_T(B)  \>\simeq\> 2\cdot\left(\ln {B^{-1}}+\eta_0\right) + H(\bB^{-1}) 
\>\simeq\>  D_1(B) + \lrang{D}\>.
\end{equation}
Here the first term is the single-jet shift, $D_1\simeq \ln(p_0/(BQ))
= \ln B^{-1}+\eta_0$; 
the second term, 
\begin{equation}
\label{eq:lrD}
\lrang{D} =  
 \frac{\pi}{2\sqrt{C_F\as}}
             +\frac{3}{4} -\frac{\beta_0}{6C_F} + \eta_0 \>,
\end{equation}
is nothing but the \NP\ 
correction to the {\em mean}\/ single-jet broadening
\eqref{eq:npmean1},
\begin{equation}
   \lrang{B}_1-\lrang{B}_1^{(\PT)} = \frac{\eps}{2}\> \lrang{D}\>. 
\end{equation}
The physical origin of this result can be simply understood.  In the
region under consideration, ($\cR'\!<\!\sqrt{\as}$), the multiplicity
of perturbative gluon radiation is small.  The non-perturbative shift
is then dominated by the fluctuations in which one of the two jets is
responsible for the perturbative component of the event broadening:
$B_L\ll B_R\approx B_T$, or vice versa.  Non-perturbative effects in
the {\em narrower}\/ jet are better pronounced: in the absence of
perturbative radiation the direction of the quark momentum stays
closer to the thrust axis thus bringing in a large contribution
$\lrang{D}\propto{1/\sqrt{\as}}$ to the shift $D_T$.  It describes the
non-perturbative correction to the {\em mean}\/ broadening of the
narrower jet.  This contribution is practically $B$ {\em
  independent}\/: its residual $B$ dependence emerges at the level of
a $\cO{\sqrt{\as}\ln^2B}$ correction.  In these circumstances the
$B$ dependence of the total shift coincides with that of a single jet.

In the opposite regime, $\as\ln^2B\!\ge\! 1$, multiple perturbative
radiation becomes a way of life.  Here both jets take responsibility
for perturbative broadening, and the coefficient of $\ln B$ in $D_T$
grows bigger than unity.  It doubles in the limit of extremely small
$B$ values, $\as\ln^2B\!\gg\! 1$ ($\cR'\!\gg\!\sqrt{\as}$), where the
differential spectrum flattens off (and then starts to decrease, when
$\cR'\!>\!1$) because of severe Sudakov suppression.  In this regime
the jets are forced to share $B$ equally, and we should expect
\begin{equation}
  \label{eq:DTlarge}
  D_T(B) \simeq 2D_1\left({B}/{2}\right)\cdot 
\left[\,1\>+\> \cO{(\as\ln^2B)^{-1}}\,\right],   
\quad   \cR'\gg\sqrt{\as}\>.  
\end{equation}
Indeed, at $\as\ln^2B\!\ge\! 1$ ($\cR'\!\ge\!\sqrt{\as}$) we have $
H(x)\simeq 1/\cR'\ll \ln B^{-1}$.  Near the maximum of the
distribution, $\cR'\ge 1$, the relative size of the $H$-contribution
becomes as small as $H/D_T\simeq (\cR')^{-1}/\ln B^{-1} = \cO{\as}$.
The first two terms in \eqref{eq:DTansb} then combine into
\eqref{eq:DTlarge}.

\section{Comparison with experimental data}
\label{sec:data}

\begin{figure}[ht]
  \begin{center}
    \epsfig{file=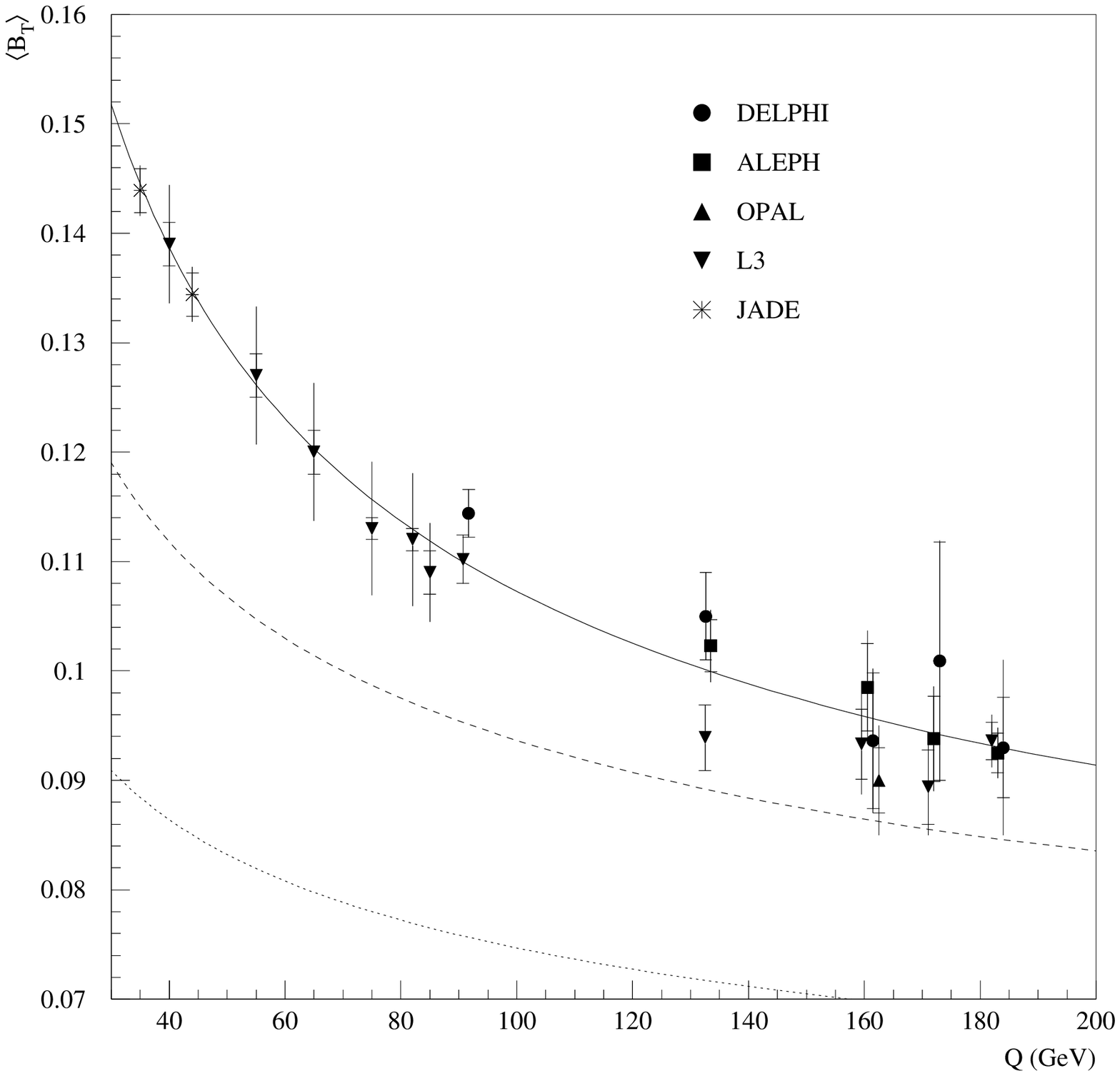,width=0.47\textwidth}
    \epsfig{file=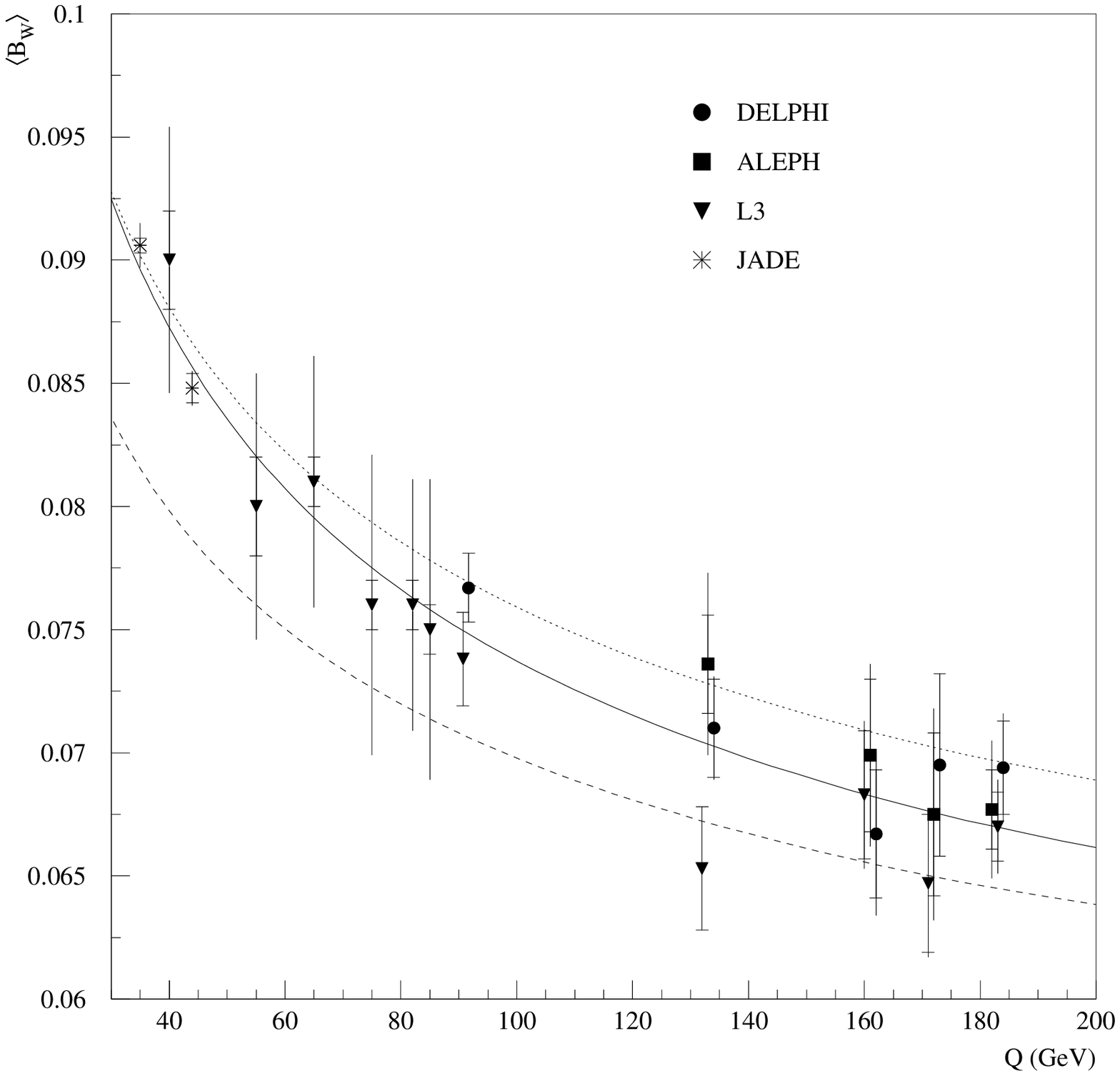,width=0.47\textwidth}
    \caption{Energy dependence of $\lrang{B}_T$ and $\lrang{B}_W$. The 
      dotted line is the LO result; the dashed line is LO$+$NLO and
      the solid line, LO$+$NLO$+$power correction, \eqref{eq:mean1Tnp} and
      \eqref{eq:meanWIDE} respectively.}  
    \label{fig:meanBt} 
  \end{center}
\end{figure}

\subsection{Means}
Using formulas (\ref{eq:collMnBT},\ref{eq:collMnBW}) we perform fits
to the data for the mean total and wide-jet broadenings and compare to
the values obtained for the thrust, the heavy-jet mass and the
$C$-parameter (with fixed-order perturbative coefficients taken from
\cite{JADE98}).

The results that we obtain are (fitting to data from
\cite{ALEPH,L3,DELPHI,DELPHIdist,OPAL,OPALdist,SLD} and references in
\cite{Bethke}, partially based on a pre-existing compilation
\cite{WickeComp}) 
\begin{center}
\begin{tabular}{|c|c|c|c|} \hline
Variable & $\as$ & $\alpha_0$ & $\chi^2$/d.o.f.\\ \hline
$B_T$ & $0.1170 \pm 0.0023$ & $0.4508 \pm 0.0225$ &  $14.9/(23-2)$ \\ \hline
$B_W$ & $0.1189 \pm 0.0025$ & $0.3911 \pm 0.0305$ &  $12.8/(22-2)$ \\ \hline
$1-T$ & $ 0.1177 \pm 0.0013$ & $0.4976 \pm 0.0087$ & $57.0/(40-2)$ \\ \hline
$C  $ & $0.1206 \pm 0.0021$ & $0.4527 \pm 0.0110$ &  $10.7/(10-2)  $ \\ \hline
$M^2_h/Q^2$ & $0.1171 \pm 0.0012$ & $0.5602 \pm 0.0224$ &
$15.2/(27-2)$ \\ \hline 
\end{tabular}
\end{center}
Figure~\ref{fig:meanBt} shows the a comparison between the data and
the fits for $B_T$ and $B_W$.

\subsection{Distributions}

We consider only the $B_T$ distribution, for the sake of illustration.
The relevant equations are (\ref{eq:collBT},\ref{eq:collDT}) for the
non-perturbative shift and the perturbative spectrum is as discussed
in section~\ref{sec:collPTspect}. We fit 8 data sets (using log-$R$
matching, with the distribution constrained to go to zero at the
kinematic limit)
\cite{OPALdist,OPAL,JADEresurrect,SLD,DELPHIdist,DELPHI} and
obtain

\begin{center}
\begin{tabular}{|c|c|c|c|} \hline
$\as$ & $\alpha_0$ & $\chi^2$/d.o.f.\\ \hline
$0.1158 \pm 0.0007$ & $0.5368 \pm 0.0077$ &  $68.7/(59-2)$ \\ \hline
\end{tabular}
\end{center}

\begin{figure}[ht]
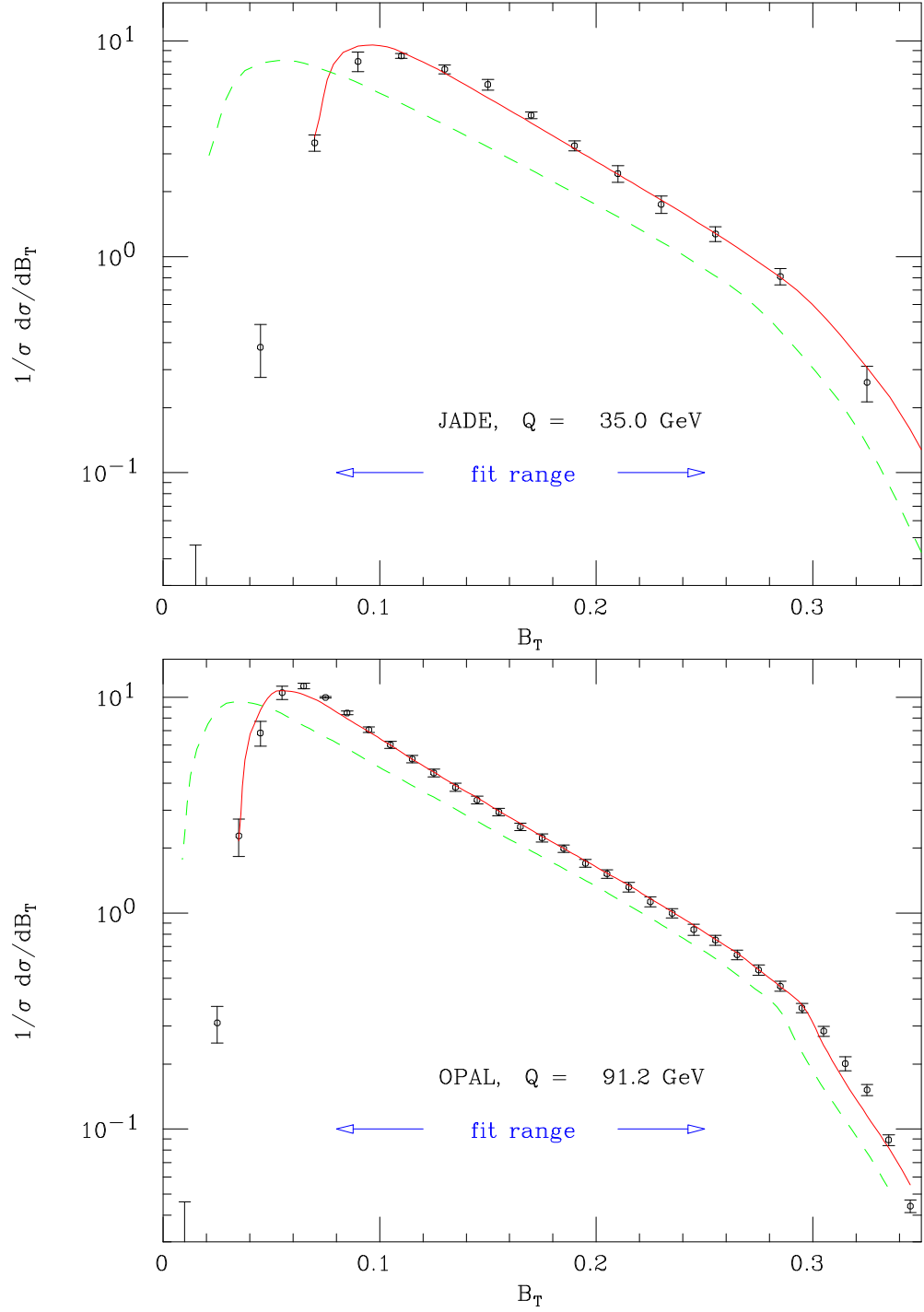

  \begin{center} 
    \epsfig{file=btdist-jade35.eps,angle=90,width=0.8\textwidth} \\
    \epsfig{file=btdist-opal91.eps,angle=90,width=0.8\textwidth}
    \caption{$B_T$ distributions compared to JADE (35 GeV)
      \cite{JADEresurrect} and OPAL (91.2 GeV) \cite{OPALdist}
       data. The dashed lines are resummed distributions without the power 
       correction, while the solid lines are the resummed predictions
       with the power correction.
  \label{fig:btdist}}
  \end{center}
\end{figure}

\noindent
Two examples of the distributions are shown in
figure~\ref{fig:btdist}, together with JADE (35~GeV) and OPAL (91.2
GeV) data.

\section{Conclusions/Discussion}
 \label{sec:Conclusions}

In this paper we have demonstrated that the sensitivity of the broadening
measure to large parton rapidities has made the non-perturbative shift
in the $B$ spectra $B$ dependent.  As a result, the \PT\ 
distribution gets shifted to larger $B$ values and {\em squeezed}\/ on
the way, in accord with the recent experimental
findings~\cite{JADE98}.  This result supersedes the previous
expectation of the $\ln Q$-enhanced shift~\cite{DLMSuniv}.
It is worthwhile  noticing that the correct result contains a single
universal \NP-parameter $\cP$ which puts broadening on an equal footing
with other jet shapes $1\!-\!T$, $C$, etc.  The previous (incorrect) result
contained an additional \NP-parameter, $Q_B$ in \eqref{old}, related
to the log-moment of the coupling defined similarly to
\eqref{eq:a0def} but with an extra factor $\ln k$ under the integral.

\clearpage

The reason why the shift in the $B$ spectra has become $B$ dependent
is the interplay between \PT\ %
and \NP\ %
gluon radiation effects.  The radiation of gluons with fixed
transverse momentum does not depend on the gluon rapidity as defined
with respect to the {\em quark}\/ direction. At the same time, the
broadening accumulates the moduli of transverse momenta of final
partons with respect to the {\em thrust axis}.  The latter, starts to
differ from the quark direction when the normal \PT\ %
radiation is taken into account.  This mismatch does not matter, at
next-to-leading accuracy, for the \PT\ %
analysis but plays a crucial r\^ole for the \NP\ %
effects both in the $B$ means and distributions.

If one naively assumes that the quark direction coincides with that of
the thrust axis then the contributions from gluons ({\em
  gluers}~\cite{gluer}) with finite transverse momenta,
$k_t\sim\LQCD$, and rapidities up to the kinematically allowed value
$\eta_i\le \eta_{\max}\simeq\ln(Q/k_{ti})$ sum up to provide the $\log
Q$-enhanced \NP\ %
shift in the $B$ spectrum, see \eqref{eq:old2}.  However, gluers with
{\em large energies}\/, $k_{0i}\ga k_{ti}/\Theta_q$ are collinear to
the {\em quark}\/ direction rather than to that of the thrust axis and
therefore do not contribute to $B$ because of the CIS nature of the
observable (gluons collinear to the quark increase $B$ by exactly the
same amount by which the quark contribution to $B$ is reduced due to
the longitudinal momentum recoil).  As a result, the \NP\ %
contribution to $B$ comes out proportional to the quark rapidity,
$\log 1/\Theta_q$, with $\Theta_q$ the quark angle with respect to the
thrust axis which is due to radiation of \PT\ %
gluons.  Since, kinematically, $\Theta_q\propto B$, averaging $\log
1/\Theta_q$ over the \PT\ %
distribution in $\Theta_q$ results in the $\log B$ enhancement of the \NP\ %
shift.

\paragraph{NP effects in the presence of PT radiation.}

Disregarding the ever-present \PT\ %
radiation is known to produce confusing results in a number of
cases.  For example, the first-order (one-gluon) analysis of the \NP\ %
effects in the heavy-jet squared mass produced the wrong expectation:
adding a single gluer to the $q\bar{q}$ system in $e^+e^-$ (as the
third and {\em only}\/ secondary parton) and constructing $M^2$ of the
$qg$ system we find a $1/Q$ confinement contribution to the squared
mass of the {\em heavy}\/ jet, the one our gluer belongs to.
Meanwhile, the opposite {\em lighter}\/ jet acquires
neither a \PT\ %
nor a \NP\ %
contribution to the mass.  As a result the \NP\ %
correction to $M^2_H$ comes out equal to that for thrust,
\begin{equation}
  \label{eq:wrongMH}
  c_T=c_{M^2_T}=c_{M^2_H}\,,\quad  c_{M^2_L}=0\>.  
\end{equation}
In reality there are always normal \PT\ 
gluons in the game which are responsible for the bulk of the jet mass:
$M^2_H/Q^2\sim \as > M^2_L/Q^2\sim \as^2 \gg \delta M^2_{\NP}$.  In
these circumstances it is the \PT\ 
radiation to determine which of the two jets is {\em heavier}.  The
gluer(s) contribute equally to both jets,
\begin{equation}
  \label{eq:rightMH}
  c_T=c_{M^2_T}=2c_{M^2_H}= 2c_{M^2_L}\>.   
\end{equation}
It is worthwhile remarking that experimental analyses carried out
before 1998 were based on the wrong expectation \eqref{eq:wrongMH}.

Another important example of the interplay between  \NP\ %
and \PT\ %
effects is given by {\em higher moments}\/ of jet shapes, e.g.
$\lrang{(1\!-\!T)^n}$.  For such a quantity one obtains, symbolically,
$$
\lrang{\left((1\!-\!T)_{\PT} +
    \frac{\LQCD}{Q}\right)^{\!\!n} } \simeq \as + \as\frac{\LQCD}{Q} +
  \ldots  \left(\frac{\LQCD}{Q}\right)^{\!\!n}, \qquad n\ge 2 .
$$
The leading power-suppressed contribution is at the level of $1/Q$.
At the same time, 
the one-gluer analysis for such an observable would formally produce 
$$
\lrang{(1\!-\!T)^n}_{\mbox{\NP\ 
        only}} \>\>\simeq\> \frac{\cA_n}{Q^n},
$$
which \NP\ 
correction is suppressed as a high power of $1/Q$.  In the presence of 
\PT\ 
radiation, however, the leading $1/Q$ contribution is still here,
though reduced by the $\as(Q)$ factor but far more important than the
$1/Q^n$ term.

Another ``mistake'' of this sort brings us closer to the $B$ issue.
Consider the transverse momentum broadening of the
current-fragmentation jet in Deep Inelastic Scattering (DIS), that is
the sum of moduli of transverse momenta of particles in the current
jet.  Adding a gluer to the Born (parton model) quark scattering
picture we get {\em three}\/ equal contributions to $B$: two
contributions from the quark $p$ which recoils against the gluer $k$
emitted either in the initial (IS) or in the final (FS) state,
$|\vec{p}_\perp|=|\vec{k}_\perp|$, and one contribution from the gluer
itself when it belongs to FS.  Taking into account the \PT\ radiation,
however, the FS quark has already got a non-zero transverse momentum,
$p_\perp^{\PT}\sim \as\cdot Q$, a substantial amount compared to
$k_\perp\sim\LQCD$.  In this environment the direct gluer's
contribution is the only one to survive: the \NP\ recoil upon the
quark gets degraded down to a $1/Q^2$ effect after the azimuthal
average is performed,
$$
\lrang{|\vec{p}_\perp|} = \lrang{|\vec{p}_\perp^{\>\PT}-\vec{k}_\perp|}
= p_\perp^{\PT} +\cO{\frac{\LQCD^2}{p_\perp^{\PT}}}.
$$
The true magnitude of the $1/Q$ contribution turns out to be a factor 
{\bf three} smaller than that extracted from the one-gluer analysis.

\paragraph{A preliminary health report.}
Now that the loophole in the theoretical treatment of the jet
broadenings has been eliminated, one can return to the much debated
issue of the universality of confinement effects in event shapes.

\begin{figure}[htbp]
  \begin{center}
    \epsfig{file=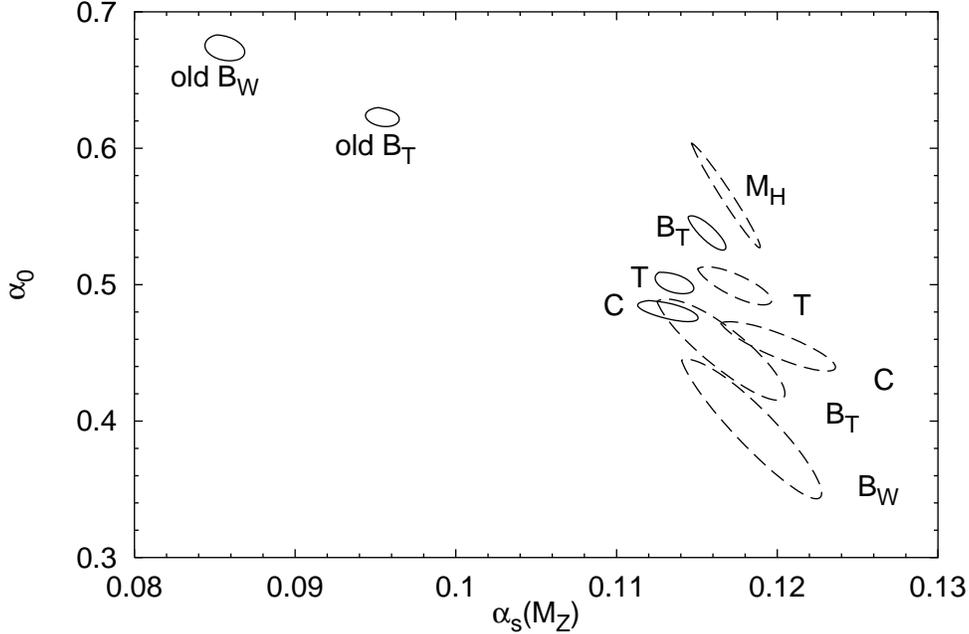}
\end{center}
\caption{95\% CL contours for jet shape
  means (dashed) and some distributions (solid). The curves for the
  $T$, $C$ and old $B_T $ and $B_W$ distributions are taken from
  \cite{JADE98}. The curves for the means are to be taken as purely
  indicative since we have not accounted for the correlations between
  systematic errors (which, where available, are added in quadrature
  to the statistical errors). Additionally for some observables we may
  not have found all the available data.}
\label{fig:CL}
\end{figure}

To illustrate the overall consistency of the universality hypothesis
we show in Figure~\ref{fig:CL} 2-standard-deviation contours in the
$\as$-$\al_0$ plane for a range of means and distributions. The curves
labelled ``old'' are the results to the fits \cite{JADE98} using the
old (wrong) formulas. The situation with the total broadening
distribution is greatly improved by the updated theoretical treatment.
We expect the wide broadening distribution to be equally improved, but
this remains to be verified.

The fits for $\as$ and $\al_0$ from the mean values are also generally
consistent with each other and with those from the distributions.
However, the agreement between different event shapes is still not
perfect. In the case of the heavy-jet mass we believe that this may in
part be related to the treatment of particle masses, which have more
effect on jet masses than on the thrust or the $C$-parameter (which
are both defined exclusively in terms of 3-momenta). We leave this
question for future consideration.

\paragraph{Outlook.} Another important issue is that of the power
correction to the jet-broadening in DIS. Formally the extension of our
results to the DIS case is quite a non-trivial exercise. As a first
step it would be necessary to carry out a resummed \PT\ %
calculation for the DIS broadening. This has so far not been done. 

The situation for the mean broadening measured with respect to the
thrust axis is fairly simple though, since (modulo factors of two
associated with the definition of the broadening in DIS
\cite{DISmilan}) it is equivalent to a single hemisphere in $e^+e^-$:
\begin{equation}
\label{eq:DISBthrust}
\lrang{B}_{\mathrm{DIS,thrust}} -
\lrang{B}_{\mathrm{DIS,thrust}}^{(\PT)}    
= \eps
\left( \frac{\pi}{2\sqrt{C_F\ascmw(\Qbar)}} +\frac{3}{4}
  -\frac{\beta_0}{6C_F} + \eta_0  + \cO{\sqrt{\as}} \right).
\end{equation}
For the mean broadening defined with respect to the photon ($z$) axis
the situation is more complicated because of the dependence on
perturbative initial-state radiation. To a first approximation, at
moderate $x$, one can view the DIS event as a \emph{rotated} $e^+e^-$
event where the broadening is measured in the right hemisphere with
respect to the
axis of the quark in the left hemisphere: i.e.\ %
the relevant transverse momentum for determining the rapidity
available to the \NP\ %
correction is $\tpt=|\vtpt_1 + \vtpt_2|$. Since this is very similar
to $\max(\tpt_1,\tpt_2)$ we have a situation like that for the
wide-jet broadening, and the leading power correction is suppressed by
factor $\sqrt{2}$ compared to \eqref{eq:DISBthrust}:
\begin{equation}
\label{eq:DISBz}
\lrang{B}_{\mathrm{DIS,}z} -
\lrang{B}_{\mathrm{DIS,}z}^{(\PT)}    
= \eps
\left( \frac{\pi}{2\sqrt{2C_F\ascmw(\Qbar)}} +\frac{3}{4}
  -\frac{\beta_0}{12C_F} + \eta_0 + \cO{1}\right) .
\end{equation}
Even though we have chosen to include some subleading terms of
$\cO{1}$, it is likely that there are other terms of $\cO{1}$, arising
through the $x$ dependence of the problem.  In particular, for the
important case of small x which is presently being studies at HERA the
process is dominated by the boson-gluon fusion mechanism and the
analogy with the two quark jets in e+e- gets lost.  Parton
multiplication in the initial state of the DIS process is more
intensive ($t$-channel gluon exchange) and increases with $\ln 1/x$
leading to increasing characteristic transverse momentum of the struck
quark, which contributes to $B_{\mathrm{DIS,}z}$.

Still more damaging will be $x$-dependent  
effects at $1-x\ll1$ where phase-space restriction on \PT %
transverse parton momenta may destroy the leading $1/\sqrt{\as}$ term
through a dependence on $\ln (1-x)$.

So \eqref{eq:DISBz} can be used only for moderate $x$, and even then
one should remember that it is subject to subleading corrections of
$\cO{1}$.

Finally let us remark that $1/Q$ power effects can be envisaged (and
should be studied) also for other jet characteristics such as shape
variables of 3-jet events (thrust minor, acoplanarity) as well as, for
example, in the distributions of the accompanying $E_T$ flow in
hadronic collisions and DIS.

\section*{Acknowledgements}

We are grateful to Siggi Bethke, Othmar Biebel, Mrinal Dasgupta, Pedro
Movilla Fernandez, Klaus Hamacher, Hasko Stenzel, Bryan Webber, Daniel Wicke
and Giulia Zanderighi for helpful discussions and suggestions.

\appendix

\section{ Perturbative and non-perturbative radiator}
 \label{Ap:PNPrad}
The two-loop radiator in the soft limit is given by (see \cite{DLMSuniv})
\begin{multline}
\label{R2loop}
\cRt(\nu,b;p)={4C_F} \int \frac{d\al}{\al} \frac{d^2\tkt}{\pi \tkt^2}
\>\left(\frac{\as(0)}{4\pi} +\chi(\tkt^2)\right)\> [1-u(k)] \\
+4C_F \int d\Gamma_2(k_1,k_2)\left(\frac{\as}{4\pi}\right)^2\frac1{2!}
M^2(k_1,k_2)
\;[1-u(k_1)u(k_2)] \>,
\end{multline}
where, for the right-broadening distribution, the source $u(k)$ is given by 
(see \eqref{uki})
\begin{equation}
  \label{probe}
1-u(k)\>=\>\left[1 - e^{-\nu(k_{t}-\al p)/Q}e^{-i\vec{b}\vtkt}\right]
\vartheta(\al-k_t/Q)\>, 
\qquad
\vec{k}_t=\vtkt +\al \vec{p}\>,
\end{equation}
with the right-hemisphere constraint \eqref{eq:rhem} included. 

The function $\chi(\tkt^2)$ is the virtual correction to one-gluon emission.
In the physical scheme which defines the coupling as the intensity 
of soft-gluon radiation~\cite{CMW}, it can be written in terms of the 
dispersive integral
\begin{equation}\label{chidef}
\chi(\tkt^2)= \int_0^\infty \frac{d\mu^2 \>\tkt^2}{\mu^2(\tkt^2+\mu^2)}
\left(\frac{\as}{4\pi}\right)^2 \left\{
-2C_A\ln\frac{\tkt^2(\tkt^2+\mu^2)}{\mu^4}
\right\} .
\end{equation}
The collinear divergence in $\chi$ is compensated by that of the real
two-parton contribution described by the matrix element $M(k_1,k_2)$. 
Taken together with an ill-defined
$\as(0)$ of one-gluon emission, they participate in forming a finite
running coupling.

The two-parton phase space in \eqref{R2loop} is
\begin{equation}
  d\Gamma_2(k_1,k_2)= dm^2\, \frac{d^2\tkt}{\pi} \frac{d\al}{\al}\>
 \cdot dz\,\frac{d\phi}{2\pi}\>,
\end{equation}
The first three variables $\al=\al_1+\al_2$,
$\vtkt=\vtkti{1}+\vtkti{2}$ and $m^2=(k_1+k_2)^2$ 
are those of the parent gluon $k$,
while $z$, $1\!-\!z$ 
and $\phi$ are the momentum fractions and the relative azimuth of the 
two secondary partons, $q\bar{q}$ or gluons:
\begin{equation}
  \label{kin}
\al_1=z\al\,,  \>   \al_2=(1-z)\al\,,\quad   
m^2=z(1-z)q_t^2\>,\quad
 \vec{q}_t = \frac{\vtkti{1}}{z} - \frac{\vtkti{2}}{1-z}\>.
\end{equation}
Hereafter we choose $\phi$ as the angle between 2-vectors $\vec{q}_t$
and $\vtkt$.

The probing functions $u(k_1)$, $u(k_2)$ depend on all the parton
variables.
In order to extract the contribution responsible for the running
coupling we split 
\begin{equation}
  \label{usplit}
 1-u(k_1)u(k_2)\>=\> \left[\,1-u(k)\,\right] + \left[\,u(k)
-u(k_1)u(k_2)\,\right] ,  
\end{equation}
where we have introduced a ``probing function'' $u(k)$ for the parent gluon. 
There are various ways to define the source $u(k)$ if $k$ is {\em massive}. 
The prescription we choose consists in substituting 
$ \tkt^2 \to \tkt^2+m^2 $ in the massless expression \eqref{probe}.
In particular, the transverse momentum with respect to the thrust
axis, $k_t$,
$$
k_t^2= \tkt^2+(\al\tpt)^2+2\al\tpt\tkt\cos\psi_{kp}\>,
$$ 
gets replaced by 
\begin{equation}
  \label{ktprime}
{k_t'}^2 \>\equiv\>\tkt^2+m^2+(\al\tpt)^2
+2\al\tpt\sqrt{\tkt^2+m^2}\cos\psi_{kp}\>. 
\end{equation}
Thus we define the inclusive probe (first term in \eqref{usplit}) by
\begin{equation}
\label{uincl}
1-u(k)\>=\>
\left(1 -
  e^{-\nu(k_{t}'-\al\tpt)/Q}e^{-ib\sqrt{\tkt^2+m^2}\cos\psi_{kb}}\right)
\cdot\vartheta(\al-k_t'/Q)\>.   
\end{equation}
The inclusive contribution depends only on the parent gluon variables,
so that the two-parton matrix element $M^2(k_1,k_2)$ can be integrated
over $z$ and $\phi$ to give (see \cite{DLMSuniv})
\begin{equation}
\label{M2int}
\int  dz\>\frac{d\phi}{2\pi} 
\frac 1{2!} M^2(k_1,k_2) 
= \frac{1}{m^2(m^2+\tkt^2)}
\left\{-\beta_0+2C_A\ln\frac{\tkt^2(\tkt^2+m^2)}{m^4}\right\},      
\end{equation}
with $ \be_0\>=\> \frac{11}{3}C_A -\frac23n_f$
the first beta function coefficient.
We treat the $\be_0$ term of this equation together with $\as(0)$ to
form the so-called {\em naive contribution}, the
logarithmic term together with $\chi(\tkt^2)$  
({\em inclusive correction}), and the second term of the trigger function 
\eqref{usplit} as the {\em non-inclusive correction}.

At the perturbative level, the naive term provides the dominant
contribution, while the inclusive and non-inclusive terms stay at the
level of the next-to-next-to-leading two-loop correction.  As far as
the non-perturbative $1/Q$ correction to the perturbative radiator is
concerned, all three give comparable contributions, the latter two
providing the so-called rescaling Milan factor to the naive one.

\paragraph{Naive contribution.} The naive contribution reads
\begin{equation}
  \label{cR0}
 \cRt_0(\nu,b;p) \> \equiv\>
4C_F \int \frac{dm^2 d\tkt^2}{\tkt^2+m^2}
\left\{\frac{\as(0)}{4\pi}\delta(m^2)
-\frac{\beta_0}{m^2}\left(\frac{\as}{4\pi}\right)^2\right\}
\cdot \Omega_0\left({\tkt^2+m^2}\right).
\end{equation}
Here we have introduced the ``naive trigger function'' representing 
the $[1-u(k)]$ factor integrated over $\al$ and $\psi_{kp}$: 
\begin{equation}
  \label{Om0}
\begin{split}
\Omega_0\left({\tkt^2+m^2}\right)
= \int_{\sqrt{\tkt^2+m^2}/Q}^1 \frac{d\al}{\al}
\int_{-\pi}^{\pi}\frac{d\psi_{kp}}{2\pi}\> \left[\,1-
 e^{-\nu(k_{t}'-\al\tpt)/Q}e^{-ib\sqrt{\tkt^2+m^2}\cos\psi_{kb}}\,\right],
\end{split}
\end{equation}
with $k_t'$ defined in \eqref{ktprime}. 
The lower limit of the logarithmic $\al$-integration, corresponding 
to separation between the two hemispheres, is actually $k_t'/Q$. 
However for $\tpt\ll Q$, this limit can be approximated by the
$\tpt$-independent value given here.

\paragraph{Inclusive correction.}
The inclusive correction can be represented in terms of a difference
between the naive trigger functions for $m\neq0$ (real) and $m=0$
(virtual contribution) as
\begin{equation}\label{cRin}
\begin{split}
\cRt_{in}(\nu,b;p) &={8C_FC_A} \int \frac{dm^2}{m^2}
\left(\frac{\as}{4\pi}\right)^2 \int \frac{d\tkt^2}{\tkt^2+m^2}
\;\ln\frac{\tkt^2(\tkt^2+m^2)}{m^4}\;
\left(\Omega_0(\tkt^2+m^2)-\Omega_0(\tkt^2)\right).
\end{split}
\end{equation}

\paragraph{Non-inclusive correction.}
Finally, the non-inclusive correction describes the mismatch between
the actual contribution to $B_R$ from two partons and that of their parent:
\begin{equation}\label{cRnonint}
\cRt_{ni}(\nu,b;p)\>=\>
4C_F\int dm^2 \>{d\tkt^2}\>\frac{d\phi}{2\pi}\> dz\>
\left(\frac{\as}{4\pi}\right)^2 
\frac1{2!} M^2(m^2,\tkt^2,z,\phi)\cdot\Om_{ni}\,,
\end{equation}
with the non-inclusive ``trigger function''
\begin{equation}
\label{nitrigger}
\Om_{ni}\>\equiv\> \int_0^1\frac{d\al}{\al}\int\frac{d\psi}{2\pi}
\;[\, u(k_1+k_2)-u(k_1)u(k_2)\,]\,.
\end{equation}

\subsection{PT contribution to the radiator \label{Ap:PNPradP}}

In the perturbative evaluation of the naive contribution we neglect
$m^2\ll\tkt^2\ll Q^2$ in the trigger function and use the dispersive
relation for $\as$,
\begin{equation}
\int_0^{Q^2}\frac{dm^2\>\tkt^2}{m^2+\tkt^2}  
\left\{{\as(0)}\delta(m^2)
-\frac{\beta_0}{m^2}\,\frac{\as^2}{4\pi}\right\}
\>=\> \as(\tkt)\cdot \left( 1+\cO{\as^2}+\cO{\frac{\tkt^2}{Q^2}}\right),
\end{equation}
to obtain
\begin{equation}
  \label{RPT}
  \cRt_0^{(\PT)}(\nu,b;p)= \frac{C_F}{\pi} \int\frac{d\tkt^2}{\tkt^2}\>
{\as(\tkt)}\cdot \Omega_0(\tkt^2)\>. 
\end{equation}
The inclusive and the non-inclusive terms do not contribute to the
\PT\ radiator at two loops. This is due to the fact that, in spite of
the singular behaviour of the matrix element, $M^2\propto1/m^2$, the
$m^2$-integrals in \eqref{cRin} and \eqref{cRnonint} converge, because
the trigger functions $\Omega_0(\tkt^2+m^2)-\Omega_0(\tkt^2)$ and
$\Om_{ni}$ vanish in the collinear parton limit.  As a result, these
contributions are determined by the non-logarithmic integration
regions $ \tkt\sim m \sim Q/\nu $ and provide negligible corrections
to the radiator of the order of $ \as^2\cdot \ln\nu
\>\Longrightarrow\> \as^2\ln B\>, $ with a single-logarithmic
enhancement factor originating from the $\al$- (rapidity-)
integration.

The \PT\  %
expression \eqref{RPT} can be greatly simplified, to our accuracy,
by observing that one can neglect the difference between quark and
thrust axis and set $p=0$ in the trigger function \eqref{Om0}.
Moreover, for large $\nu$ and $b$ one can substitute the trigger
function $\Omega_0$ by a suitably chosen cutoff in the phase space
integration.

\paragraph{Negligible $\tpt\neq0$ effect.}
As we shall see later, the effects of quark recoil ($\tpt\ne0$)
modifies the structure of the answer for the {\em non-perturbative}\/
contribution.  However, at the {\em perturbative}]\/ level a small
departure of the quark direction from that of the thrust axis proves
to be negligible, producing a $\cO{\as^2\ln B}$ subleading correction
which lies beyond the scope of the approximation, i.e.\ %
keeping under control terms $\cO{\as^n\ln^m B}$ with $m\ge
n$~\cite{CTW,DLMSbroad}.

To verify this we observe that the difference between the \PT\ 
radiator \eqref{RPT}, \eqref{Om0} and its $\tpt\!=\!0$ value,  
\begin{equation}
\label{jk2}
  \cRt^{(\PT)}(b,\nu;0)= \frac{C_F}{\pi}
 \int_0^{Q^2} \frac{d\tkt^2}{\tkt^2}\>
 \as(\tkt)\> \ln\frac{Q}{\tkt}
\left[\,1-e^{-\nu \tkt/Q}J_0(b\tkt)\,\right],  
\end{equation}
in the large-$\nu$ limit is a function of the ratios $\nu\tpt/Q$ 
and $bQ/\nu$, 
\begin{equation}
  \label{eq:17}
   \cRt^{(\PT)}(b,\nu;\tpt)- \cRt^{(\PT)}(b,\nu;0)  = 
 f\left(\frac{\nu\tpt}{Q}, \frac{bQ}{\nu}\right) \>+\>\cO{\frac{1}{\nu}}\>.
\end{equation}
To verify this we introduce the rescaled dimensionless variables
$q=\frac{\nu k_t}{\al Q}$ and $ \vec{u}= \vb Q/\nu$ to write
$$ 
\cRt^{\PT}(\tpt)- \cRt^{\PT}(0)  = 
\frac{C_F}{2\pi}\int_0^1 \frac{d\al}{\al}\>\as\int_0^{\nu }\frac{d^2q}{\pi} 
\left\{ 
\frac{1-e^{-\al (q-\nu\tpt/Q)}e^{-i\al\vec{u}(\vec{q}-\nu\vtpt/Q)}
}{(\vec{q}-\nu\vtpt/Q)^2}
-\frac{1-e^{-\al q}e^{-i\al\vec{u}\vec{q}}}{q^2}
\right\}.
$$
Eq.~\eqref{eq:17} follows from the fact that the $q$-integral
converges and can be extended to run up to $q=\infty$ instead of
$q=\nu$.  As a result, it does dot depend explicitly on $\nu$ except
through the combinations $\nu\tpt/Q$ and $bQ/\nu$.  In the essential
region, $bQ\sim \nu\sim Q/\tpt\gg1$, these combinations are of the
order one, so that
\[
f\left(\frac{\nu\tpt}{Q}, \frac{bQ}{\nu}\right) = \as(Q)\cdot \cO{1} \>.
\]

\paragraph{Radiator as a function of a single variable.}
The final simplification of \eqref{jk2} for large $\nu$ and $b$ is 
obtained by the replacement~\cite{DLMSbroad}
\begin{equation}
  \label{mu}
\left[\,1-e^{-\nu k_t/Q}J_0(bk_t)\,\right] \Rightarrow \Theta(k_t-Q/\bmu)\>,
\qquad \bmu \>\equiv e^{\gamma_E} \>\frac{\nu+\sqrt{\nu^2+(bQ)^2}}{2}\>,
\end{equation}
which introduces a negligible error $\cO{\as}$ to the radiator and
gives, in the next-to-leading logarithmic approximation, 
\begin{equation}
 \cR^{(\PT)}(\bmu)\>=\>\frac{2C_F}{\pi}\int_{Q/\bmu}^Q\frac{dk_t}{k_t}\>
 \as(k_t)\> \left( \ln\frac{Q}{k_t} -\frac 34\right)\>.
\end{equation}
Here we have replaced the soft $d\al/\al$ factor in the trigger
function \eqref{Om0} by the exact $q\to qg$ splitting probability to
account for a ``hard'' subleading correction originating from the
region $\al\sim 1$. The result is the $-3/4$ term in the integrand.
While necessary for achieving single-logarithmic accuracy in the {\em
  perturbative}\/ treatment, the hard part of the kernel is
irrelevant, as we shall see below, for the {\em non-perturbative}\/
contribution which is dominated by the region $\al\sim\tkt\sim m\ll
Q$.

We give expressions for the radiator for three forms of the coupling:

\noindent $\bullet$ For fixed coupling
\begin{equation}
   \cR^{(\PT)}(\bmu) = \frac{C_F\as}{\pi} \left(\ell^{\,2}- \th\ell\right),
   \quad \ell\equiv \ln\bmu\>.
\end{equation}
$\bullet$ Taking into account the one-loop running of $\as$ we obtain
\begin{equation} 
\label{as1}
 \cR^{(\PT)}(\bmu)
= \frac{4C_F}{\beta_0} \int_0^{\ell} d\ell'\> \frac{\ell'- \frac34}{L-\ell'}
= -\frac{4C_F}{\beta_0}\left[\,\left(L-\frac34\right)
\ln\left(1-\frac{\ell}{L}\right)+\ell\,\right], 
\quad L\equiv\ln\frac{Q}{\Lambda}.
\end{equation}
$\bullet$ To evaluate the radiator with the two-loop running coupling we
use 
\begin{equation}
  \label{as2}
  \as(Q)\>=\>\frac{2\pi}{\beta_0(L+\gamma \ln L)}\>, \qquad 
\gamma= \frac{\beta_1}{\beta_0^2}\,,\>\>\beta_1=51-\frac{19}{3}n_f\>,
\end{equation}
to obtain 
\begin{equation}
  \label{cR2}
  \cR^{(\PT)}(\bmu)\>=\>\frac{4C_F}{\beta_0}\left\{
   \left(T-\tq\right)\ln\frac{T}{T_{\bmu}}-
   (T-T_{\bmu})-\half\gam\ln^2\frac{T}{T_{\bmu}}
   \right\}.
\end{equation}
Here
\begin{equation}
 \label{t}
  T\>\equiv\>\frac{2\pi}{\beta_0\>\as(Q)}\>+\>\gamma\>,
\qquad 
  T_{\bmu}\>\equiv\>\frac{2\pi}{\beta_0\>\as(Q/\bmu)}\>+\>\gamma\>,
\end{equation}
and $\as$ is the running coupling in the so-called physical
scheme~\cite{CMW} which can be related to the standard $\MSbar$ coupling
by \eqref{eq:CMWvsMSbar}. 

The radiators with the two-loop and one-loop $\as$ deviate at the level 
of a $\cO{\as^3\ln^3\bmu}$ term which contribution is under control and
should be kept in the \PT\ %
distributions. At the same time, for the sake of simplicity we 
use the (two-loop) radiator with the one-loop coupling \eqref{as1}
for evaluating the \NP\ %
contributions to the means and spectra. For example, in the means we
know that $\beta_0$ appears as a constant term, suppressed with
respect to the leading term by a factor of $\sqrt{\as}$, see
\eqref{eq:npmean1}.  It is clear therefore that a $\beta_1$
contribution will enter at the level of a term of $\cO{\as}$, which is
beyond our control.

In conclusion, the \PT\ 
part of the Mellin transform of the single-jet broadening in
\eqref{Sigma} reads
\begin{equation}
  \label{sigmaPT}
  \sigma^{(\PT)}(\nu)= \int_0^{\infty} bdb\>e^{-\cR^{(\PT)}(\bmu)}
\int_0^{\infty} \tpt d\tpt\> e^{-\nu \tpt/Q}  J_0(b\tpt)\>.
\end{equation}
The $\tpt$-integration gives 
\begin{equation}
\label{eq:ptint}
  \int_0^\infty \frac{\tpt d\tpt}{Q^2}\>e^{-\nu \tpt/Q}\>J_0(b\tpt)\>=\>
 \frac{1}{\nu^2y^3}\>,
\qquad y\>\equiv\>\frac{\sqrt{\nu^2+(bQ)^2}}{\nu}\>,
\end{equation}
which results in
\begin{equation}
\label{eq:Sigpty}
   \sigma^{(\PT)}(\nu)= \int_1^{\infty} \frac{dy}{y^2}\>
 e^{-\cR^{(\PT)}(\bmu)} \>; \qquad \bmu=\half e^{\gamma_E}(1+y)\nu\,.
\end{equation}

\subsection{Non-perturbative contribution of the radiator \label{Ap:PNPradN}}
To extract the power-suppressed contribution to the radiator we 
use the procedure developed in~\cite{DMW}.
It is based on introducing the effective coupling $\ae(m^2)$ related
to the usual coupling constant $\as$ via the dispersive relation
\begin{equation}
  \label{disp}
\frac{\as(k)}{k^2}\>=\> \int_0^\infty dm^2\> \frac{\ae(m^2)}{(m^2+k^2)^2}\>.
\end{equation}
We then substitute the non-perturbative ``effective coupling modification'' 
$\delta\ae$ for $\ae$ and look for the leading non-analytic in $m^2$ term 
in the $m^2\to0$ limit.

To obtain the \NP\
contribution to the radiator we substitute
\begin{equation}
\left\{{\as(0)}\delta(m^2) -\frac{\beta_0}{m^2}\,\frac{\as^2}{4\pi}\>
+\cdots\right\}\cdot \>=\>  \ae(m^2)\>\frac{-d}{d m^2}\cdot 
\end{equation}
in the naive, inclusive and non-inclusive contributions.

\paragraph{Naive contribution.}  We obtain
\begin{equation}
 \label{cR0eff}
\begin{split}
\cRt_0(\nu,b;p)
&= \frac{C_F}{\pi}\int_0^{Q^2} dm^2\> \ae(m^2)
\left(\frac{-d}{dm^2}\right) \int_0^{Q^2} 
\frac{d\tkt^2}{\tkt^2+m^2}\>\Omega_0(\tkt^2+m^2)\\
&=\frac{C_F}{\pi}\int_0^{Q^2} \frac{dm^2}{m^2}\> \ae(m^2)\>
 \Omega_0(m^2)\>+\> \cO{\as(Q)}.
\end{split} 
\end{equation} 
At the \PT\ %
level, this expression is equivalent to \eqref{RPT}, with the
non-logarithmic perturbative $\as(Q)$ correction coming from the
region $m^2\sim Q^2$.

To trigger the leading power correction in \eqref{cR0eff} 
we substitute $\delta\ae$ for $\ae$ and consider the leading 
non-analytic in $m^2$ term $\Omega(m^2)\propto \sqrt{m^2}$
which is obtained by linearising $\Omega$ in $m\sim\tkt, Q \al\ll Q$. 
In this approximation the \NP\ 
component of the trigger function, $\delta\Omega$, does not depend on
$b$, and we get
\begin{equation}
  \label{eq:11}
\delta\Omega_0(m^2) = \nu\frac{\sqrt{m^2}}{Q}\cdot
\int_{\tpt/Q}^{{\tpt}/{\sqrt{m^2}}}
\frac{du}{u}\>  \int_0^{2\pi}\frac{d\psi}{2\pi}
\left(\sqrt{1+u^2+2u\cos\psi}-u\right) \>,
\end{equation}
where we have introduced $u=\al\tpt/\sqrt{m^2}$ as an integration
variable.  The $u$-integral converges and is determined by the region
$\tpt/Q\!<\!u\!\la\! 1$.  Therefore, in the $m^2\to0$ limit we can
replace the actual upper limit, $\tpt/{\sqrt{m^2}}\>\gg\>1$, by
$\infty$ (neglecting the $\cO{m^2/\tpt}$ contribution to
$\delta\Omega$, which is analytic in $m^2$ and thus does not produce a
\NP\ correction). We have then
\begin{equation}
  \label{eq:12}
\delta\Omega_0(m^2) \>=\> \frac{\sqrt{m^2}}{Q}\cdot\rho(\tpt)\>, \qquad
\rho\>\equiv\> \nu \left(\ln\frac{p_0}{\tpt}+\cO{\frac{\tpt}{Q}}\right),
\end{equation}
where the integration constant $p_0/Q$ is given by
$$ 
\ln \frac{p_0}Q =
\int_{0}^{\infty}\frac{du}{u}\>\int_0^{2\pi}\frac{d\psi}{2\pi} 
\left[\,(\sqrt{1+u^2+2u\cos\psi}-u)-\vartheta(1-u)\,\right]
= -0.6137056 \equiv \eta_0\,.
$$ 

Recalling the definition of the first non-analytic moment of
$\delta\ae$,
$$
 A_1 \>=\> \frac{C_F}{2\pi}\int_0^\infty \frac{dm^2}{m^2}\cdot m\> 
\delta\ae(m^2)\>,
$$
we finally obtain
\begin{equation}
\label{deltacR0}
  \cR_0^{(\NP)}(\nu,b;p)\>=\> \frac{C_F}{\pi} \int
  \frac{dm^2}{m^2}\>\delta\ae(m^2) \cdot \delta\Omega(m^2) 
\>=\> \nu\cdot \frac{2A_1}{Q}\,\ln\frac{p_0}{\tpt}\>. 
\end{equation}

\paragraph{Milan factor.}
It is straightforward to verify that the inclusive and non-inclusive
trigger functions, in the linear approximation in $\tkt\sim m$, 
are proportional to the same function $\rho(\tpt)$ that determines the 
naive trigger function $\delta\Omega_0$ given in \eqref{eq:12}.
We have 
\begin{eqnarray}
\delta\Omega_{in} &=& \rho\cdot
\left(\sqrt{\tkt^2+m^2}-\tkt\right), 
\\
\delta\Omega_{ni} &=& \rho\cdot \left( \tkti{1}+\tkti{2} 
  - \sqrt{\tkt^2+m^2}\right). 
\end{eqnarray}
Such a structure is typical for $1/Q$ power corrections to various 
jet shapes and leads to the {\em universal}\/ rescaling of the naive
contribution \eqref{deltacR0} 
by the so-called Milan factor, for details see~\cite{DLMSuniv}.

With account of the Milan factor, the full two-loop \NP\ 
component of the broadening radiator reads
\begin{equation}
 \cR^{(\NP)} \>=\>   \cR_0^{(\NP)}\cdot\cM
 \>=\> \nu\cdot \eps \> \ln\frac{p_0}{\tpt}\>,
\qquad
\eps = \frac{2A_1\cM}{Q}\>.
\end{equation}
In conclusion, the Mellin transform of the single-jet broadening in
\eqref{Sig} reads
\begin{equation}
  \label{fineapp1}
  \sigma(\nu)= \int_0^{\infty} bdb\>e^{-\cR(\bmu)}
\int_0^{\infty} \tpt d\tpt\> e^{-\nu \tpt/Q}  J_0(b\tpt)
\left(\frac{\tpt}{p_0}\right)^{\nu\eps}\>.
\end{equation}
We need to evaluate the $\tpt$, $b$ and $\nu$ integrals.  The result
of the $\tpt$ integration reads
\begin{equation}
\label{eq:nptint}
  \int_0^\infty pdp\>e^{-\nu \tpt/Q}
  \left(\frac{p}{p_0}\right)^{\nu\eps}\>J_0(bp)\>=\>
 \frac{I_{\eps}(\nu,y)}{\nu^2y^3}\>,
\qquad y\>\equiv\>\frac{\sqrt{\nu^2+(bQ)^2}}{\nu}\>,
\end{equation}
where $I_{\eps}(\nu,y)$ is related to Legendre function and has the
following small-$\eps$ expansion,
\begin{equation}
\label{I}
\begin{split}
  I_{\eps}(\nu,y) \>&=\>\Gamma(2+\nu\eps)\left(\nu
    y\frac{p_0}{Q}\right)^{-\nu\eps}\>
  yP_{1+\nu\eps}\left(\frac1y\right) \\
&=\>1+\nu\eps\left(2-\gam_E-\eta_0 +\ln\frac{1+y}{2y^2}-\ln\nu-y\right)
  +\cO{\eps^2}\,; \quad \eta_0\equiv \ln\frac{p_0}{Q}\,.
\end{split}
\end{equation}

\section{ Calculation of $f(0)$ \label{Ap:f(0)}}
To calculate the non-perturbative correction to $\lrang{B}_1
=\half\lrang{B}_T$ we need to find the value of $f$ in the origin.
To this end we split $f(\nu)$ given in \eqref{eq:fdef} into two pieces:
\begin{equation}
\begin{split}
f(\nu)&= f_1(\nu) + f_2(\nu) \\
f_1(\nu)&=  \int_0^\infty \frac{\nu\>b\,db}{(b^2+\nu^2)^{3/2}} \left(2
-\gamma_E -\eta_0 +\ln\frac{\nu+\sqrt{b^2+\nu^2}}{2(b^2+\nu^2)}
\right)e^{-\cR(\bmu)} , \\
f_2(\nu) &=  -  \int_0^\infty \frac{b\,db}{b^2+\nu^2}e^{-\cR(\bmu)}.
\end{split}
\end{equation}
Here and in the rest of this section, for simplicity, we measure $b$
in units of $Q$.

After extracting $-\ln\nu$ from the first term, the remaining integral 
in $y={\sqrt{b^2+\nu^2}/\nu}$ converges and is therefore 
determined by the region $b\sim\nu\to0$ where the radiator can be
dropped as $\cO{\as}$. 
We get
\begin{equation}
  f_1(\nu\to0) \>=\> 2-\gamma_E-\eta_0 +\rho(0) -\ln2\nu \>+\> \cO{\as}
 \>\simeq\> \ln{2}-\eta_0 -\gamma_E-\ln\nu \>. 
\end{equation}
The second piece which is logarithmic in $b$ we integrate by parts to write
\begin{equation}
\begin{split}
 f_2(\nu\to0) \>&=\> -\half\left. \ln\left(b^2+\nu^2\right)e^{-\cR(\bmu)}
\right|^{b=\infty}_{b=0} + \half\int_0^\infty d\left(e^{-\cR(\bmu)}\right)
\ln\left(b^2+\nu^2\right) \\
&= \ln\nu+\int_0^\infty d\left(e^{-\cR(\bmu)}\right)\ln b\>,
\qquad
\bmu\to \half e^{\gamma_E}b\>,\quad\mbox{with}\>\> \nu\to0\>.
\end{split}
\end{equation}
In the region of finite $b=\cO{1}$ the true radiator is $\cO{\as}$ and
its contribution to the integral there can be disregarded. To make it
more explicit we cut the integral from below at some value $b>b_0\sim
1$ and integrate by parts back again:
$$
 \int_{b_0}^\infty d\left(e^{-\cR(e^{\gamma_E}b/2)}\right)\ln b
 = -\ln b_0\cdot (1+\cO{\as}) - \int_{b_0}^\infty \frac{db}{b}
\>e^{-\cR(e^{\gamma_E}b/2)}\>.
$$
The answer to the required accuracy does not depend on $b_0$, which
can therefore be chosen arbitrarily.
It is convenient to take $b_0=2 e^{-\gamma_E}$ in order to represent 
\begin{equation}
 f_2(\nu\to0) \>=\> \ln\nu  -(\ln2-\gamma_E) \>- \int_{1}^\infty \frac{dz}{z}
\>e^{-\cR(z)}\>,
\end{equation}
which finally gives 
\begin{equation}
\label{eq:f0}  
f(0) \>=\> -\eta_0 \>-\int_{1}^\infty \frac{dz}{z} \>e^{-\cR(z)}
\>=\>-\eta_0 -E(1)\>.
\end{equation}

\subsection{Mean straight from the distribution (consistency check)}

We used the residue of $\sigma(\nu)/\nu^2$ at $\nu=0$ to calculate
$\lrang{B}$  
according to \eqref{eq:B1mean}. Here we verify that the same answer
follows, within our accuracy, from integrating the spectrum:  
\begin{equation}
\begin{split}
  \lrang{B}_1&= \int_0^{B_m} dB\>B\>\frac{d}{dB}\left(\Sigma_1(B)-1\right)
= -\int_0^{B_m} dB \left(\Sigma_1(B)-1\right)\\
&= \lrang{B}_1^{\PT} -\int_0^{B_m} dB
\left(\Sigma_1(B)-\Sigma_1^{(\PT)}(B)\right) . 
\end{split}
\end{equation}
The $\NP$ component of the mean can be obtained directly by
integrating  \eqref{eq:Rop} over $B$ up to the kinematical boundary
$B_m\sim1$: 
\begin{equation}
\begin{split}
\lrang{B}_1 -\lrang{B}_1^{(\PT)} &= -\frac{\eps}{2}
e^{-\cR(e^{-\partial_a})}
\frac{C(a)-\psi(1+a)+\ln(2B_m)-a^{-1}}{\Gamma(1+a)}\>
\left(\frac{2B_m}{e^{\gamma_E}\lambda }\right)^{a} . 
\end{split}
\end{equation}
Setting $a\sim\cR'\sim\as$ to zero in the non-singular pieces we get 
\begin{equation}
\lrang{B}_1 -\lrang{B}_1^{(\PT)}  \>=\> -\frac{\eps}{2} 
\left[\, \ln {B_m} -\eta_0 -e^{-\cR(e^{-\partial_a})} \frac1{a}\cdot 
\frac{(B_m)^a\left(\frac{2}{\lambda(a) }\right)^{a}
}{\Gamma(1+a)e^{a\gamma_E}} \right] .
\end{equation}
In the final term, singular in $a$, the last factor should be expanded
to first order in $a$.  Using
$$ 
   \left(\frac2{\lambda(a)}\right)^a = 1+\cO{a^2}\>, \quad 
   \Gamma(1+a)e^{a\gamma_E} = 1 + \cO{a^2}\>,
$$
we observe that the dependence on the kinematical boundary value 
$B_m$ cancels,
$$
e^{-\cR(e^{-\partial_a})} \frac{(B_m)^a}{a} \simeq \ln B_m 
+ e^{-\cR(e^{-\partial_a})} \frac{1}{a}\>, 
$$
and the result given in \eqref{eq:npmean1} is reproduced.

\subsection{Calculation of $E(x)$
\label{Ap:E(x)}}

We define the function
\begin{equation}
\begin{split}
  \label{eq:EdefAP}
  E(x)\>&=\> \left. e^{\cR(x)}\, e^{-\cR(xe^{-\partial_a})}a^{-1}
     \right|_{a=0}\>.
\end{split}
\end{equation}
Replacing 
\begin{equation}
  \label{eq:replace}
 \frac1a \>=\> \int_1^\infty \frac{dy}{y}\> y^{-a}\>,
\end{equation}
we apply the rule \eqref{eq:reg} to absorb the $y^{-a}$ factor into a
rescaling of the argument of $\cR$, get rid of the differential
operator and represent $E$ in terms of the logarithmic integral,
\begin{equation}
\label{eq:Edef2} 
E(x)\> =\>  e^{\cR(x)} \int_1^\infty \frac{dy}{y}\> e^{-\cR(xy)}
\>=\>  \int_x^\infty \frac{dz}{z}\> e^{\cR(x)-\cR(z)} \>. 
\end{equation}
The perturbative radiator (as a function of a single variable $x>1$)
is defined in \eqref{eq:cRone}.
The first coefficients of the logarithmic expansion of the
\PT-radiator with the one-loop $\as$ ($\ell=\ln x$, $L=\ln
Q/\Lambda$) are
\begin{eqnarray}
  \label{eq:Rexpan}
  \cR'(x) &=& \>\>\frac{4C_F}{\beta_0}\frac{\ell-\tq}{L-\ell} 
      \>\>  = \quad \frac{2C_F\as(Q/x)}{\pi}\,\left(\ln x-\tq\right) \>, \\
  \cR''(x) &=&   \frac{4C_F}{\beta_0}\frac{L-\tq}{(L-\ell)^2} 
      = \quad \frac{2C_F\as(Q/x)}{\pi}\cdot\frac{\as(Q/x)}{\as(\bar{Q})}\>, \\
  \cR'''(x) &=&  \frac{8C_F}{\beta_0}\frac{L-\tq}{(L-\ell)^3} 
        = \quad \frac{\beta_0}{2C_F}\, \left(\cR''\right)^2\cdot 
       \frac{\as(\bar{Q})}{\as(Q/x)} \>; \\
\label{eq:Rn}
\cR^{(n)}(x) &=& \left(2\cR''(x)\right)^{\frac{n}{2}} \cdot \frac{(n-1)!}{2}\>
  \left(\frac{\be_0^2\,\as(\bar{Q})}{16\pi C_F}\right)^{\frac{n-2}{2}}\>.
\end{eqnarray}
with $\bar{Q}=Qe^{-\tq}$.
We also point out the structure of the combination
\begin{equation}
  \label{eq:R1R2}
  s=\frac{\cR'(x)}{\sqrt{2\cR''(x)}}\>=\>
  \sqrt{\frac{C_F\as(\bar{Q})}{\pi}}\, \left(\ln x-\tq\right).
\end{equation}
The next step is to substitute $(\ln x \!-\!\partial_a)$ for $\ln x$ 
to construct the operator
\begin{equation}
\cR(x) - \cR(xe^{-\partial_a}) 
\>=\> \cR'(x)\partial_a -\half\cR''(x)\partial_a^2
                        + \os\cR'''(x)\partial_a^3 \>+\> \ldots
\end{equation}
We start by noticing that
\begin{equation}
F(a;\cR'') =  \exp\left\{ -\half \cR''\, \partial_a^2 \right\} a^{-1}
  \>=\> \sqrt{\frac{\pi}{2\cR''}}\, N\left(\frac{a}{\sqrt{2\cR''}}\right), 
\end{equation}
where the function $N$ is related with the probability integral,  
\begin{equation}
  \label{eq:Ndef}
  N(t)= \frac{2e^{t^2}}{\sqrt{\pi}}
\int_t^\infty dx\> e^{-x^2}\>=\>e^{t^2}(1-\Phi(t))
= \frac2{\sqrt{\pi}}e^{t^2}\mathop{\rm Erfc}(t)\>,
\end{equation} 
and has the following behaviour:
\begin{eqnarray}
\label{eq:Ntailorsm}
N(t)&=&1-\frac{2t}{\sqrt{\pi}} + t^2 - \frac{4t^3}{3\sqrt\pi}
+ \ldots\>, \qquad t\ll 1\>, \\
\label{Ntailorlg}
N(t)&=& \frac{1}{\sqrt{\pi}\,t}\left(1-\frac{1}{2t^2} +\frac{3}{t^4} + 
  \ldots \right) , \qquad t\gg 1\>.
\end{eqnarray}
As a result,
$$
 F(a;\cR'') \>\simeq\> {a}^{-1}\qquad \mbox{for}\>\> a\gg\sqrt{\cR''}\>.
$$
Now we introduce the first derivative to obtain
\begin{equation} 
\begin{split}
\exp\left\{ \cR'\partial_a -\half \cR''\, \partial_a^2 \right\} a^{-1}
&= e^{\cR'\partial_a} \, F(a;\cR'') \>=\>  F(a+\cR';\cR'') \\
&=\sqrt{\frac{\pi}{2\cR''}}\, N\left(\frac{a+\cR'}{\sqrt{2\cR''}}\right). 
\end{split}
\end{equation}
To estimate contributions of higher derivatives, $n\ge3$,  
we use \eqref{eq:Rn} to derive
\begin{equation}
  \frac{\cR^{(n)}\left(\partial_a\right)^n}{n!} F(a+\cR';\cR'')
= \frac{\be_0\sqrt{\pi}}{16C_F\,n}\>
\frac{\as(\bar{Q})}{\as(Q/x)}\> 
\left(\frac{\be_0^2\,\as(\bar{Q})}{16\pi C_F} \right)^{\frac{n-3}{2}}
\>\left.\frac{d^n N(t)}{dt^n}
\right|_{t=s+\frac{a}{\sqrt{2\cR''}}} \!\!\!.
\end{equation}
We conclude that $n\!=\!3$ contributes at the level of $\cO{1}$, while 
the contributions of higher derivatives are down by the factor 
$(\sqrt{\as})^{n-3}\ll1$. 

Evaluating the third derivative, 
\begin{equation}
 \exp\left\{ \frac{\cR'''\partial_a^3}{6}\right\} F(a+\cR';\cR'')
 \>\simeq\> \left(1+\frac{\cR'''\partial_a^3}{6}\right)  F(a+\cR';\cR'')\,,
\end{equation}
we obtain
\begin{equation}
\left. \frac{\cR'''\partial_a^3}{6}\>  F(a+\cR';\cR'')\right|_{a=0}
 = -\frac{\cR'''}{3(\cR'')^2}\, X(s)
= - \frac{\be_0}{6C_F}\frac{\as(\bar{Q})}{\as(Q/x)}\,X(s)\>,  
\end{equation}
where the function $X$ has been introduced,
\begin{equation}
  X(t) = -\frac{\sqrt{\pi}}{8} \frac{d^3N(t)}{dt^3}
\>=\> \int_0^\infty zdz\> e^{-z-2t\sqrt{z}}\>.
\end{equation}
It has the asymptotic behaviour
\begin{eqnarray}
\label{eq:Xtailorsm}
X(t) &=&  1- \frac{3\sqrt{\pi}}{2}\,t + 4t^2 \>+\> \cO{t^3}\>, \quad
t\ll 1\>; \\
\label{eq:Xtailorlg}
X(t) &=& \frac{3}{4\,t^4} \>+\> \cO{t^{-6}}\>, \quad
t\gg 1\>.
\end{eqnarray} 
For \eqref{eq:EdefAP} we finally obtain
\begin{equation}
  \label{eq:derfin}
E(x) 
= \sqrt{\frac{\pi}{2\cR''}}\, N\left(\frac{\cR'}{\sqrt{2\cR''}}\right)
- \frac{\be_0}{6C_F}\,X\left(\frac{\cR'}{\sqrt{2\cR''}}\right) \>+\> \ldots
\end{equation}
The neglected terms in \eqref{eq:derfin} are of relative order
${\as}$.

For $s\equiv \cR'/\sqrt{2\cR''}\ll1$ (see \eqref{eq:R1R2}) we
substitute for $N$ and $X$ the expansions \eqref{eq:Ntailorsm},
\eqref{eq:Xtailorsm} and obtain, keeping contributions up to
$\cO{\sqrt{\as}}$,
\begin{equation}
\begin{split}
  E(x) &= \sqrt{\frac{\pi}{2\cR''}}\left( 1- \frac{2s}{\sqrt{\pi}}
 + s^2 + \ldots\right) 
- \frac{\be_0}{6C_F}\left( 1- \frac{3\sqrt{\pi}}{2}\,s + \ldots\right) \\
&\simeq \frac{\pi\sqrt{\as(\bar{Q})}}{2\sqrt{C_F}\as(Q/x)}
 -(\ln x-\tq) + \frac{\sqrt{C_F\as}}{2}(\ln x-\tq)^2
- \frac{\be_0}{6C_F} + \frac{\be_0}{4C_F}\sqrt{C_F\as}(\ln x-\tq) \\
&= 
 \frac{\sqrt{\as(\bar{Q})}}{2\sqrt{C_F}} 
\left[\frac{\pi}{\as(Q/x)}+\frac{\be_0}{2}(\ln x-\tq)\right] 
-\ln x+\tq - \frac{\be_0}{6C_F} 
+ \frac{\sqrt{C_F\as}}{2}(\ln x-\tq)^2 \>.
\end{split}
\end{equation}
We get 
\begin{equation}
\label{eq:Exans}
 E(x) \>=\> 
\frac{\pi}{2\sqrt{C_F\as(
\bar
{Q})}} -\ln x+\frac34 - \frac{\be_0}{6C_F} 
\>+\> \cO{\sqrt{\as}\ln^2x}\>.
\end{equation}
For $x=1$ (the operator entering the expressions for mean broadening) 
this gives
\begin{equation}
  E(1) 
 \>=\> \frac{\pi}{2\sqrt{C_F\as(\bar{Q})}} +\frac{3}{4}
 -\frac{\beta_0}{6C_F} \>+\> \cO{\sqrt{\as}}\,.
\end{equation}

\section{Calculation of $\delta$}
 \label{Ap:delta}
Here we calculate the Mellin integral necessary to determine
$\lrang{B}_W$, 
\begin{equation}
 \delta 
\>=\> \frac12 \int_{-i\infty}^{i\infty} 
  \frac{d\nu}{2\pi i\nu} 
\left[\, f(\nu)\sigma^{(\PT)}(-\nu)- f(-\nu)\sigma^{(\PT)}(\nu)\,\right].
\end{equation}
Invoking the operator representations \eqref{eq:sigPTnuop} and
\eqref{eq:fnuoper} for $\sigma^{(\PT)}(\nu)$ and $f(\nu)$
correspondingly, we obtain
\begin{equation*}
\begin{split}
\delta &= 
e^{-\cR(z e^{-\partial_a})} e^{-\cR(z e^{-\partial_b})}
\int_{-i\infty}^{i\infty} \frac{d\nu}{4\pi i\nu}
\left(F(a)-F(b)\right)
\left(\nu\frac{\lambda(a)}{2}\right)^{-a} 
\left(-\nu\frac{\lambda(b)}{2}\right)^{-b},
\end{split}
\end{equation*}
with $z=2e^{\gamma_E}$  and
$$
F(a) = C(a)+\partial_a -\frac1a\>.
$$
It is implied that we have to set $a=b=0$ after applying the
differentiations.  Now we introduce the real integration variable
$v=-i\nu$, add the negative and positive $v$-beams into $2\Im$, and
start the $v$-integration from some finite value $v_0=\cO{1}$ so as to
ensure applicability of the large-$\nu$ logarithmic expression for the
\PT\ %
radiator.  This gives
\begin{equation*}
\delta = 
e^{-\cR(z e^{-\partial_a})} e^{-\cR(z e^{-\partial_b})}
\int_{v_0}^{\infty} \frac{dv}{2\pi v}
\left(F(a)-F(b)\right)  v^{-(a+b)}\> \sin\frac{\pi(b-a)}{2} 
\left(\frac{\lambda(a)}{2}\right)^{-a} 
\left(\frac{\lambda(b)}{2}\right)^{-b}.
\end{equation*}
Observing that the regular pieces $C(a)$, $C(b)$ cancel in the
difference $F(a)-F(b)$ at the level of $\cO{\as}$, and that the ratios
$\lambda/2$ produce negligible corrections $\cO{a^2+b^2}$, we are left
with
\begin{equation*}
\begin{split}
\delta &= \left. e^{-\cR(z e^{-\partial_a})} e^{-\cR(z e^{-\partial_b})}
 \left\{ \partial_a - \partial_b -\frac1a+\frac1b\right\}
\frac{\left(v_0\right)^{-(a+b)}}{a+b}\>
\frac{\sin\frac{\pi}{2}(b-a)} {2\pi}\right|_{a=b=0}\\
&= \left. e^{-\cR(v_0z e^{-\partial_a})} e^{-\cR(v_0z e^{-\partial_b})}
\frac{1}{a+b} \left\{ \partial_a - \partial_b -\frac1a+\frac1b\right\}
\left( \frac{b-a}{4}+\cO{(a-b)^3}\right)\right|_{a=b=0}\>,
\end{split}
\end{equation*}
where we have absorbed the power of $v_0$ into additional rescaling of 
the arguments of the \PT\ %
radiators $\cR$.  Evaluating the derivatives and using the
$a\leftrightarrow b$ symmetry we arrive at
\begin{equation*}
\delta 
=\left. \half e^{-\cR(v_0z e^{-\partial_a})} e^{-\cR(v_0z e^{-\partial_b})}
\left( \frac1{a+b}-\frac1{a}\right) \right|_{a=b=0} 
= \left. \half\left(e^{-2\cdot\cR(v_0\lambda e^{-\partial_a})} 
- e^{-\cR(v_0\lambda e^{-\partial_a})}\right) a^{-1} \right|_{a=0}.
\end{equation*}
The finite rescaling of the argument of the $\cR$ operator by the
factor $v_0z$ produces, according to \eqref{eq:Exans},  
a subleading correction $\ln(v_0z)=\cO{1}$ which is of the
relative order $\sqrt{\as}$ and should be kept under control. 
These subleading corrections however are identical for the two terms and
cancel in the difference, thus ensuring independence 
of the result on the arbitrary parameter $v_0$, at the $\cO{\as}$ level.
We conclude,  
\begin{equation}
\label{eq:numinnuans}
\delta \>=\> \left. \half\left(e^{-2\cdot\cR(e^{-\partial_a})} 
- e^{-\cR(e^{-\partial_a})}\right) a^{-1}\right|_{a=0}.  
\end{equation}
According to \eqref{eq:npcompW}, to obtain $\lrang{B}_W$ we have to add
to \eqref{eq:numinnuans} half of the \NP\ %
correction to single-jet $\lrang{B}$, that is $-\half f(0)$ with
$f(0)$ given in \eqref{eq:f0}.  The main piece of the latter cancels
the subtraction contribution described by the single-jet operator
applied to $1/a$, and we finally arrive at
\begin{equation}
\begin{split}
 \lrang{B}_W - \lrang{B}_W^{(\PT)}  &=\>  \half{\eps} \left( \left. 
e^{-2\cdot\cR(e^{-\partial_a})}  a^{-1} \right|_{a=0}
+\eta_0 \right) \\
\>&=\> \half\eps \left( \frac{\pi}{2\sqrt{2C_F\as(\bar{Q})}} 
+\frac{3}{4} -\frac{\beta_0}{12C_F} + \eta_0\right).
\end{split}
\end{equation}
This result can be obtained from that for a single-jet (total)
broadening by the simple substitution $C_F\to 2C_F$ in
\eqref{eq:Exans}.

\section{Analysis of $D_T$ \label{Ap:DT}}

Consider the singular piece of \eqref{eq:STfact},
\begin{equation}
S=e^{-\cR(xe^{-\partial_a})}e^{-\cR(xe^{-\partial_b})} \>
\> \frac{\left(\frac{\lambda(\cR')}{\lambda(a)}\right)^{a}
\left(\frac{\lambda(\cR')}{\lambda(b)}\right)^{b}}{\Gamma(1+a+b)}
\cdot\left(-\frac{b}{a}\right).
\end{equation}
Using \eqref{eq:replace} we trade the $1/a$ factor for the logarithmic
integral of the exponent of the \PT\ radiator with the rescaled
argument, as we did in \eqref{eq:Edef2}, to obtain
\begin{equation}
\begin{split}
S &= -\cR'(x)\,e^{-\cR(x)}  \int_1^\infty \frac{dy}{y}\>  e^{-\cR(xy)} 
\frac{\left(\frac{\lambda(\cR'(x))}{\lambda(\cR'(xy))}\right)^{\cR'(xy)}}
{\Gamma(1+\cR'(x)+\cR'(xy))} \>[\,1+\cO{\as}\,]\\
&\simeq -\cR'\,e^{-\cR(x)}  \int_x^\infty \frac{dz}{z}\> e^{-\cR(z)} 
\frac{\left(\frac{\lambda(\cR')}{\lambda(\cR'(z))}\right)^{\cR'(z)}}
{\Gamma(1+\cR'+\cR'(z))}\>, \quad \cR'\equiv \cR'(x)\,.
\end{split}
\end{equation}
The integrand is monotonically decreasing with $z$. 
Extracting the \PT\ %
distribution, we write
\begin{equation}
\label{eq:Hdef}
  S = -\cR'\,\frac{e^{-2\cR(x)}}{\Gamma(1+2\cR')} \cdot H(x)\>, \quad
H(x)= \int_x^\infty \frac{dz}{z}\>
e^{\cR(x)-\cR(z)}\frac{\Gamma(1+2\cR')}  
{\Gamma(1+\cR'+\cR'(z))}\>,
\end{equation}
where we have dropped the factor containing the ratio of $\lambda$
functions as it produces a negligible effect.  Taken together with the
regular piece in \eqref{eq:STfact} this leads to the expression for
the shift reported in \eqref{eq:DTans}.

For  $\cR'\ll 1$ we can expand the ratio of the $\Gamma$ functions in
\eqref{eq:Hdef} to get
\begin{equation}
\label{eq:Hans}
\begin{split}
  H(x) &=  \int_x^\infty \frac{dz}{z} \> e^{\cR(x)-\cR(z)} 
\left[\, 1 - \gamma_E\cR' \>+\>\gamma_E\cR'(z) \>+\>\ldots\, \right]
\\ 
&= \left(1-\gamma_E\cR'\right)\cdot E(x) + \gamma_E\>+\> \cO{\as
  E(x)}\,. 
\end{split}
\end{equation}
The answer for $H(\bB^{-1})$ depends on the relation between $\cR'$ and
$\sqrt{\cR''}\sim \sqrt{\as}$.

For  $s\ll1$ ($\cR'\ll\sqrt{\as}$) \eqref{eq:Hans} gives
\begin{multline}
\label{eq:Hxsmall}
  H(\bB^{-1}) \>=\>  E(\bB^{-1}) + \gamma_E + \cO{s} \>\simeq\> 
 \left( \ln \bB + \frac{\pi}{2\sqrt{C_F\as}} +\frac{3}{4}
   -\frac{\beta_0}{6C_F} 
\right) + \gamma_E \\
\simeq\> \ln B + \frac{\pi}{2\sqrt{C_F\as}} +\frac{3}{4}
-\frac{\beta_0}{6C_F}\>, 
\end{multline}
where we have substituted, according to \eqref{eq:SigmaPTfin}, 
$\bB \simeq{B}e^{-\gamma_E}$ (for small $\cR'$ one has 
$\lambda(\cR')/2=1 + \cO{\cR'}$).

\noindent
For $s\gg1$ ($\cR'\gg\sqrt{\as}$) we have instead
$$
 H(\bB^{-1}) \>\simeq\> {\cR'}^{-1}\>,
$$
which contribution to the shift becomes negligible, 
$$
\frac{H}{|D_T|} \sim \frac 1{\cR'\cdot |\ln B+\mbox{const}|} 
\sim \frac1{\as\ln^2 B} \sim s^{-2} \ll 1\>.
$$
For $\cR'\gg1$, the contribution  of $H$ is even smaller, 
$H(\bB^{-1})\ll (\cR')^{-1}$, due to additional fall-off of the
$\Gamma$-function in the integrand of \eqref{eq:Hdef}.

We conclude that the approximate expression \eqref{eq:Hans},
\begin{equation}
   H(\bB^{-1}) 
\>\simeq\> \left(1-\gamma_E\cR'\right) E(\bB^{-1}) + \gamma_E\>,
\end{equation}
with $\bB$ given in \eqref{eq:SigmaPTfin}, can be used everywhere, for
arbitrary $\ln B$ values.

\section{Numerical cross-checks and illustrations}

In this Appendix we present a numerical approach which allows us to
illustrate some important properties of the results and to estimate
(some) next-to-next-to leading effects not accounted for by the
approximate analytical expressions derived in the paper.

We express the factorised $n$-soft-parton cross section through an
equation governing a branching process, which can be implemented as a
Monte Carlo ``event'' generator --- where by event one means an
ensemble of soft gluons. For the accuracy to which we consider the jet 
broadening, it will be sufficient to consider just the emissions of
gluons from the quark and antiquark, ignoring their subsequent
branching \cite{DLMSbroad}.

Considering for the time being only a single jet, if the last gluon
was produced with transverse momentum $k_{i-1}$, the probability
distribution for the next gluon's transverse momentum ($k_{t,i}
<k_{t,i-1}$) is given by
\begin{equation}
  k_{t,i} \frac{dP}{dk_{t,i}} = 2\asb 
  \left(\ln \frac{Q}{k_{t,i}} -  \frac{3}{4}\right) 
  \Delta(k_{t,i-1},k_{t,i})\,,
\end{equation}
where $\asb= \as\cf/\pi$ and $\Delta$ resums the virtual corrections:
\begin{equation}
  \ln \Delta(k_{t,i-1},k_{t,i}) =
  -\int_{k_{t,i}}^{k_{t,i-1}} \frac{dk_t}{k_t} 2\asb
  \left(\ln \frac{Q}{k_{t}} -  \frac{3}{4}\right)\,.
\end{equation}
Technically, the branching starts from a ``fake'' gluon, $i=0$, with
\begin{equation}
  k_{t,0} = Q e^{-3/4}\,,
\end{equation}
which does not contribute to any of the momentum sums.
The azimuthal direction of each $k_{t,i}$ is chosen at random.
The quark recoil, $p_t$ is
\begin{equation}
  p_t = \left| \sum_{i=1} \vkti{,i} \right|\,,
\end{equation}
and the broadening, as in \eqref{eq:Bdef}, is
\begin{equation}
  B = \frac{1}{2Q}\left(p_t + \sum_{i=1} \left| \vkti{,i}
    \right|\right)\,. 
\end{equation}
Implementing the branching as a Monte Carlo generator makes the the
vector sum extremely simple --- and correspondingly so the analysis of
the mean $\ln p_t$ as a function of $B$, or also averaged over B.

If one uses $\as(k_{t,i})$ in the CMW, or ``physical'' scheme
\cite{CMW}, then apart from the overall normalisation, the
accuracy of the resulting description is the same as that of the
resummed expressions of \cite{CTW,DLMSbroad}, namely
next-to-leading-logarithmic in the exponent.

In practice, when comparing with the analytical results given in the
main part of the paper it is most informative to consider the results
in the limit of fixed $\as$ (i.e.\ 
setting $\beta_0=0$) so as to avoid having to introduce an arbitrary
infrared cutoff in the numerical calculation.

\begin{figure}[htbp]
  \begin{center}
    \epsfig{file=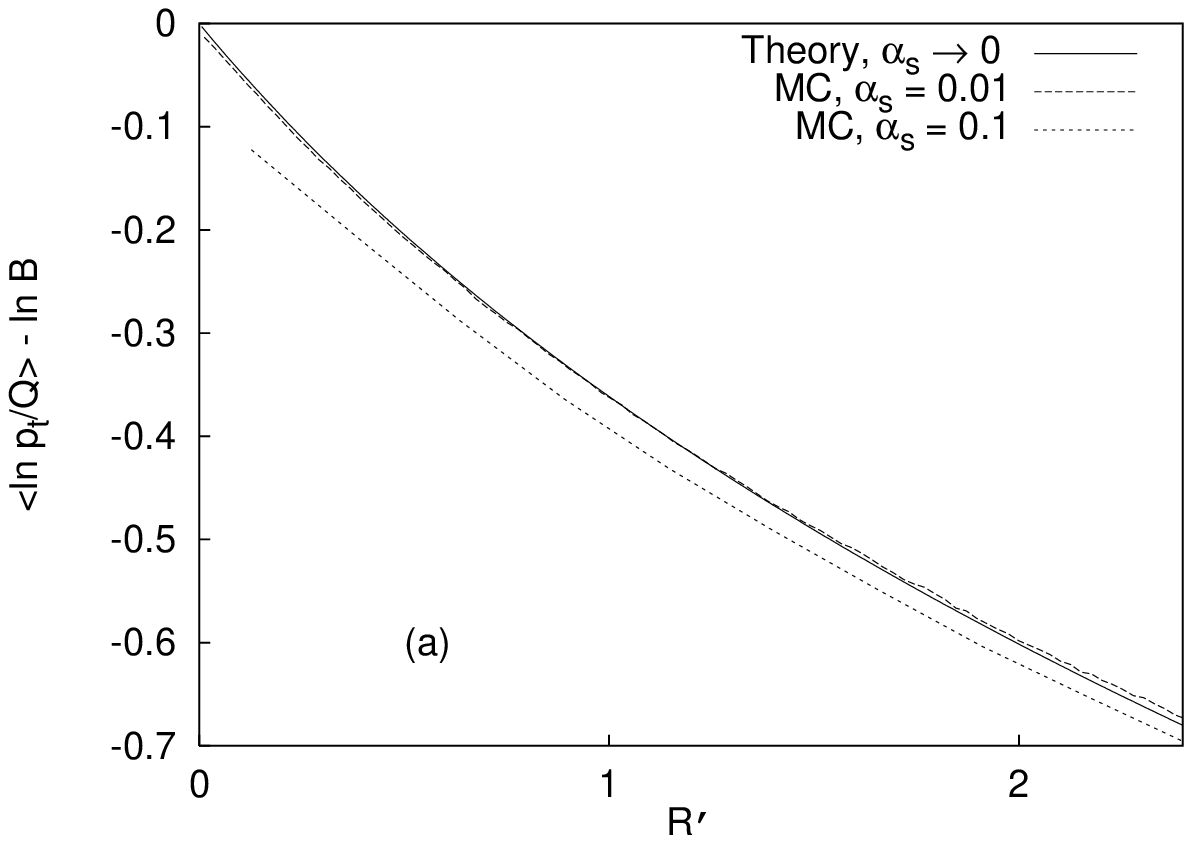}
    \epsfig{file=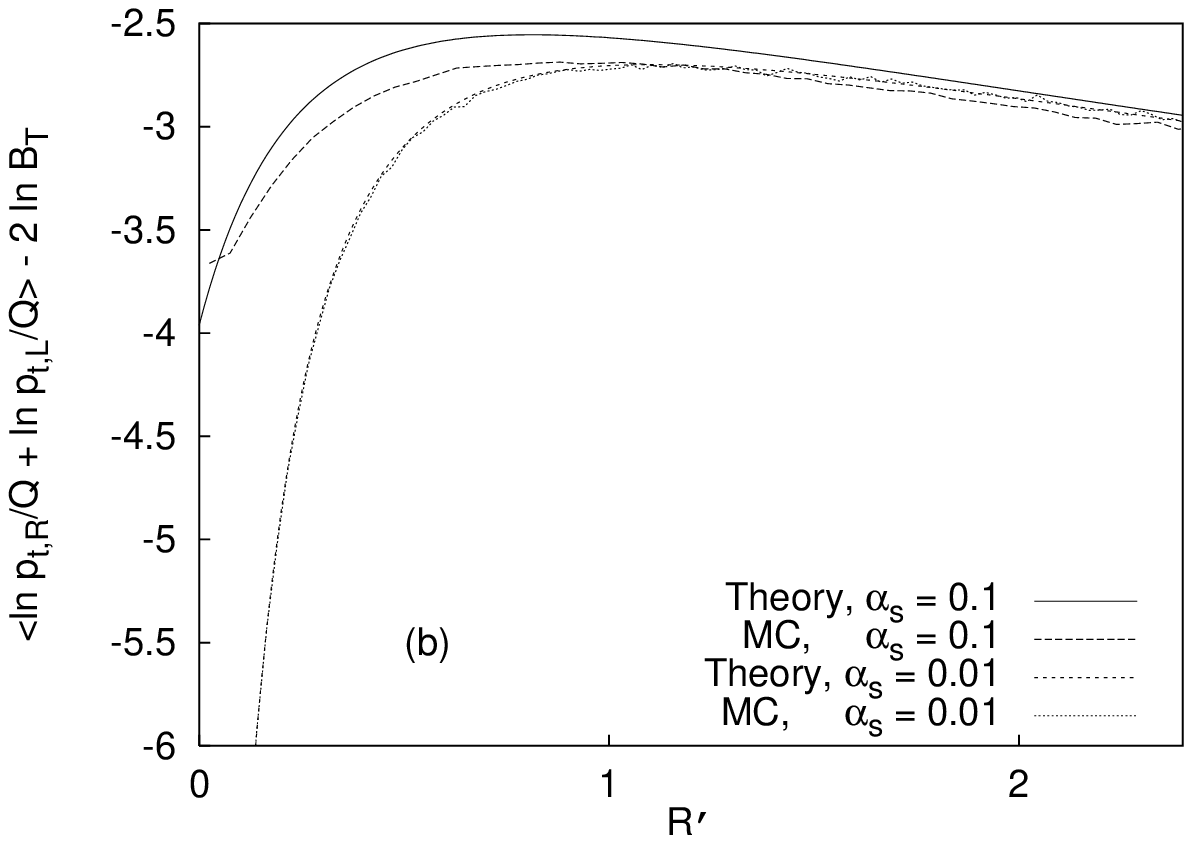}
    \caption{(a) $\lrang{\ln p_t/Q}-\ln B_1$ as a function of
      $\cR'$. The Monte Carlo results are shown for various values  
      of $\as$ while the analytic curve is independent of $\as$; (b)
      $\lrang{\ln p_{t,R}/Q + \ln p_{t,L}/Q }-2\ln B_T$ from the Monte
      Carlo and from \eqref{eq:DTans}. The Monte Carlo and theory curves
      for $\as=0.01$ lie on top of each other.}  
    \label{fig:lnptB}
  \end{center}
\end{figure}

We recall that the power correction to the broadenings depends
linearly on the the logarithm of the quark transverse momentum in each
jet. Accordingly, from \eqref{eq:D1ans} we expect that $\lrang{\ln
  p_{t}/Q} - \ln B$ in a single jet with broadening $B$, should be a
function only of $R'(B)$, and we test this by plotting $\lrang{\ln
  p_t/Q} - \ln B$ versus $R'$. The resulting curves should then be
independent of $\as$. Figure~\ref{fig:lnptB}a shows such curves for
two values of $\as$, compared with the theoretical result. In the
Monte Carlo results, there is a small dependence on the value of $\as$
--- corresponding roughly to a shift of the curves by an amount of
order $\as$. This is a sublogarithmic effect and so beyond the
accuracy of the analytical calculations (and in any case beyond the
predictive value of the Monte Carlo). For small values of $\as$ there
is good agreement between the Monte Carlo and the analytical results.

When considering the total jet broadening, $B_T$, we just generate two
independent single-jet configurations \eqref{eq:fact}, and construct
the combination $\lrang{\ln p_{t,R}/Q+\ln p_{t,L}/Q} - 2 \ln B_T$
which is a function of $R'(B_T)$ only for large values of $R'$. For
small $R'$, it goes as $1/\sqrt{\as}$. Accordingly, in
figure~\ref{fig:lnptB}b the curves for different values of $\as$
differ at smaller $R'$.  However the analytic predictions and the
Monte Carlo results remain in good agreement, except roughly at the
level of a shift of order $\as$ as in the case of the single jet
curves.

\begin{figure}[htbp]
  \begin{center}
    \epsfig{file=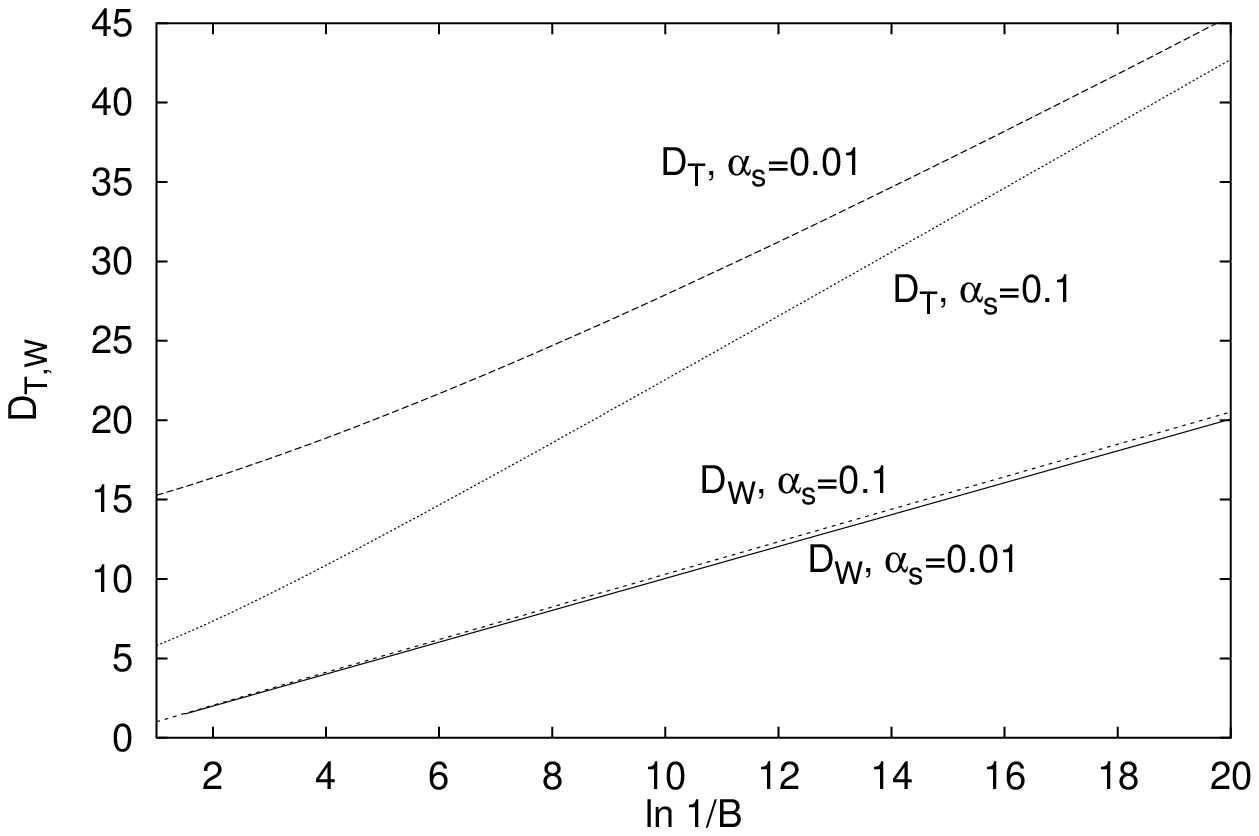}
    \caption{$D_T$ and $D_W$ as a function of $B$, for two
      values of a ``fixed'' $\as$.}
    \label{fig:btbw}
  \end{center}
\end{figure}

While the above figures demonstrate the agreement between the Monte
Carlo and the analytical results, they do not illustrate the shifts
themselves as a function of $B$. For the case of a fixed coupling,
this is done in figure~\ref{fig:btbw}. The main features are
the following: $D_W$ is practically independent of $\as$ and almost
equals $\ln 1/B$. For very small $B$, $D_T$ is practically twice $\ln
1/B$. For larger $B$ one can see that the slope of the $D_T$ curves
tends to that of the $D_W$ curves. Finally at large $B$ one sees an
offset in $D_T$ which increases as $1/\sqrt{\as}$.

\section{Collection of final formulas }
\label{sec:Collection}
We collect here for convenience the final expressions for the
broadening distributions and means, which include $1/Q$ confinement
effects and were used for the phenomenological analysis presented in
section~\ref{sec:data}.

\subsection{Expressions for the shifts}

\paragraph{The integrated wide-jet broadening distribution:} 
\begin{equation}
\label{eq:collBW}
\sigma^{-1}\int_0^B dB_W\> \frac{d\sigma}{dB_W} 
\>\equiv\> \Sigma_W(B) \>=\>  \Sigma_W^{(\PT)}(B-\half\cP D_1(B))
\end{equation}
with the $B$-dependent shift $D_1$ given by\footnote{
We remark that in $\cR'$ the $3/4$ is beyond the accuracy that we
control, however we choose to keep it because it has a clear origin,
and is among the next-to-leading corrections to the power-suppressed
contribution.}
\begin{equation}
\label{eq:collDW}
\begin{split}
D_1(B) \>&=\> \ln B^{-1} + \eta_0
-2 - \rho(\cR')+\chi(\cR')+ \psi(1+\cR')\!-\!\psi(1)\,, \\
\cR' \>&=\> 2C_F\frac{\as(BQ)}{\pi}\left(\ln B^{-1} -
  \tq\right)\>,\quad \eta_0= -0.6137056\>,
\end{split}
\end{equation}
and $\psi(z)$ the derivative of the logarithm of $\Gamma(z)$. 
The functions $\rho$ and $\chi$ are
\begin{equation}
\begin{split}
  \rho(a) &= \int_0^1 dz\> 
\left(\frac{1+z}{2z\,\lam(a)}\right)^{-a}  \ln z(1+z) \>, \qquad
  \chi(a) = \frac2a \Big(\left[\lam(a)\right]^a-1\Big), \\
& [\lam(a)]^{-a}\>\equiv\> \int_0^1 dz\>\left(\frac{1+z}{2z}\right)^{-a}.
\end{split}
\end{equation}

\paragraph{The integrated total broadening distribution:} 
\begin{equation}
\label{eq:collBT}
\sigma^{-1}\int_0^B dB_T\> \frac{d\sigma}{dB_T} 
\>\equiv\> \Sigma_T(B) \>=\>  \Sigma_T^{(\PT)}(B-\half\cP D_T(B))
\end{equation}
with the $B$-dependent shift $D_T$ given by
\begin{equation}
\label{eq:collDT}
\begin{split}
  D_T(B) \>&=\> 2D_1(B) +2[\psi(1+2\cR')-\psi(1+\cR')]\>+\> H(\bB^{-1})\>,\\
 H(x)\>&=\> \int_x^{z_0} \frac{dz}{z}\>
 e^{\cR(x)-\cR(z)}\frac{\Gamma(1+2\cR')}  
{\Gamma(1+\cR'+\cR'(z))}\>,\qquad \bB =
\frac{2B}{e^{\gamma_E}\lambda(\cR')}\,, 
\end{split}
\end{equation}
where $z_0$ corresponds to the position of the Landau pole in the
perturbative radiator $\cR(z)$, where the integrand vanishes. The form
that we use for $\cR$ is the two-loop radiator with the one-loop
coupling, which, in the physical ($\CMW$) scheme has the simple
expression
\begin{equation} 
 \cR(x)=  -\frac{4C_F}{\beta_0}\left[\,\left(L-\frac34\right)
\ln\left(1-\frac{\ln x}{L}\right)+\ln x\,\right] 
\end{equation}
where $L=\ln Q/\Lambda= 2\pi/(\beta_0 \ascmw(Q))$, and the physical
coupling $\ascmw$ is related to the standard $\alpha_{\MSbar}$ by
\begin{equation}
\label{eq:CMWvsMSbar}
  \ascmw = \alpha_{\MSbar} \left( 1 + K
    \frac{\alpha_{\MSbar}}{2\pi}\right) ,
\end{equation}
with
\begin{equation}
  \label{eq:Kdef}
   K\equiv
  \ca\left(\frac{67}{18}-\frac{\pi^2}{6}\right)-\frac{5}{9}n_f \>,
  \qquad \beta_0=\frac{11\ca}{3}-\frac{2n_f}{3}\>.
\end{equation}
The two-loop radiator with the two-loop coupling is given in
Appendix~\ref{Ap:PNPradP} in \eqref{as2}\footnote{For enthusiasts only!}.

\subsection{PT spectra}
\label{sec:collPTspect}

The resummed expressions \eqref{eq:SigmasPT}
that we derived for the \PT\ %
spectra are applicable in the small-$B$ region, and have
next-to-leading-logarithmic accuracy. For the purposes of
phenomenology, it is necessary to extend the domain of validity of the
perturbative spectra towards larger values of $B$, where it is
sufficient to use a fixed calculation. Procedures (the $R$ and
$\log$-$R$ matching schemes) for combining the fixed order and
resummed results are explained in detail in \cite{CTTW} for the thrust
and heavy-jet mass distributions, and are directly applicable also to
the case of the broadenings. The requirement for the use of these
procedures (in particular $R$-matching) is that the resummed
perturbative answer have only the following terms
\begin{equation}
  \ln \Sigma(V) = C_1 \as + \sum_{n=1}^\infty G_{n,n+1} \as^n \ln^{n+1} V
  + \sum_{n=1}^\infty G_{n,n} \as^n \ln^{n} V\,,
\end{equation}
and it mustn't have terms such as $\as^2 \ln V$. Suitable expressions for 
the broadenings were presented in \cite{CTW}, equations (18--22). As
was shown in \cite{DLMSbroad} these answers have to be modified by an
additional factor (both for the wide and total broadenings):
\begin{equation}
  \Sigma^{(\PT)}(B) = \left(\frac2{\lambda(\cR')}\right)^{2\cR'}\cdot
  \Sigma^{(\PT)}_{\mathrm{CTW}}(B)\,.
\end{equation}
It is vital that $\cR'$ here be taken as:
\begin{equation}
  \label{RpMatching}
  \cR' = \frac{2\as(Q) \cf}{\pi} \frac{\ln(1/B)}{1 -
    \frac{\alpha_s(Q)\beta_0}{2\pi}  
    \ln(1/B)}\,.
\end{equation}
We stress that the $3/4$ in $\cR'$ is a next-to-next-to-leading
effect, and as such is taken care of by the matching
procedure\footnote{We note that the $\beta_0$ that we use in this
  paper differs from that in \cite{CTW} by a factor of $4\pi$.}. For
that procedure to remain intact, it must not be included in
\eqref{RpMatching}.

\subsection{Means}

The leading power correction to the mean total broadening is 
\begin{equation}
\label{eq:collMnBT}
\lrang{B}_T - \lrang{B}_T^{(\PT)} 
= \eps
\left( \frac{\pi}{2\sqrt{C_F\ascmw(\Qbar)}} +\frac{3}{4}
  -\frac{\beta_0}{6C_F} + \eta_0  + \cO{\sqrt{\as}} \right).
\end{equation}
and the correction to the mean wide-jet
broadening is
\begin{equation}
\label{eq:collMnBW}
 \lrang{B}_W - \lrang{B}_W^{(\PT)} \>=\> \frac{\eps}{2}
\left( \frac{\pi}{2\sqrt{2C_F\ascmw(\Qbar)}} +\frac{3}{4}
  -\frac{\beta_0}{12C_F} + \eta_0    + \cO{\sqrt{\as}} \right).
\end{equation}
Here, $\Qbar = Q e^{-3/4}$. The use of $\Qbar$ rather than $Q$ as the
scale for $\as$, and the choice of $\ascmw$ rather than $\al_{\MSbar}$
both affect the results at the level of a $\cO{\sqrt{\as}}$ term,
which formally we do not control. However we prefer to keep these
corrections since they have clear physical origins (the $e^{-3/4}$
factor in the scale has about a $5\%$ effect on the fitted value of
$\al_0$, while the change from $\CMW$ to $\MSbar$ schemes has much
less effect).

\subsection{Non-PT parameter}
In order to accurately define the non-perturbative  parameter $\eps$
the problem of {\em merging}\/ the \PT\ %
and \NP\ %
contributions should be addressed. The relevant procedure was
discussed in detail in \cite{DLMSuniv}.  It includes introducing an
infrared matching scale $\mu_I$ (typically chosen to be $\mu_I=$2 GeV)
and the non-perturbative $\mu_I$-dependent phenomenological parameter
$\al_0$~\eqref{eq:a0def} which quantifies the intensity of QCD
interaction over the infrared domain, $ k\le\mu_I$.

An explicit expression for $\eps$ depends on the order to which the
perturbative contribution is computed, as well as on the scheme.  At
two-loop level, in the $\MSbar$ scheme, we have
\begin{equation}
  \label{eq:cPfin}
 \eps\> \equiv\>  \frac{4C_F}{\pi^2}\cM \frac{\mu_I}{Q}
\left\{ \alpha_0(\mu_I)- \as
  -\beta_0\frac{\as^2}{2\pi}\left(\ln\frac{Q}{\mu_I} 
+\frac{K}{\be_0}+1\right) \right\}\>;  \quad \as\equiv \al_{\MSbar}(Q)\>.
\end{equation}
The term proportional to $K$ accounts for mismatch between the
$\MSbar$ and the physical scheme, with $K$ given above in
\eqref{eq:Kdef}.  $\cM$ in \eqref{eq:cPfin} is the Milan factor
resulting from the two-loop analysis discussed in
Appendix~\ref{Ap:PNPradN} (see also~\cite{DLMSthrust}).  This factor
is universal for all $1/Q$ jet observables considered in $e^+e^-$
annihilation~\cite{DLMSuniv} and DIS processes~\cite{DISmilan} and
reads\footnote{The value $\cM \simeq 1.795$ presented in the original
  version of \cite{DLMSthrust} and used in the original version of
  this paper was wrong. The figures shown elsewhere in this paper are
  still based on the old value of $\cM$, but change only slightly with
  the new, corrected, value.}
\begin{equation}\label{cMdef}
\begin{split}
\cM \>=\>
1+{\be_0^{-1}}\left( 1.575C_A\>-\> 0.104n_f\right) 
\>=\> 1.490\>(1.430) \quad \mbox{for}\>\> n_f=3\>(0)\>.
\end{split}
\end{equation}
The perturbative terms in \eqref{eq:cPfin} 
(proportional to ${\as}$ and ${\as^2}$) 
represent the start of the series responsible for
subtracting off the infrared renormalon divergence in the {\em
perturbative}\/ contribution to the observable.

\end{document}